\documentstyle[aaspp4]{article}

\begin{document}
\title{The range of masses and periods explored by radial velocity searches 
for planetary companions}

\author{A. F. Nelson}
\affil{Department of Physics, The University of Arizona, Tucson AZ 85721}
\affil{email: andy@as.arizona.edu}
\author{J. R. P. Angel}
\affil{Steward Observatory, The University of Arizona, Tucson AZ 85721}
\affil{email: rangel@as.arizona.edu}

\begin{abstract}
 
Radial velocity measurements have proven a powerful tool for finding 
planets in short period orbits around other stars.  In this paper we 
develop an analytical expression relating the sensitivity to a periodic
signal to the duration and accuracy of a given set of data.  The effects 
of windowing are explored, and also the sensitivity to periods longer 
than the total length of observations.  We show that current observations 
are not yet long or accurate enough to make unambiguous detection of 
planets with the same mass and period as Jupiter.  However, if 
measurements are continued at the current best levels of accuracy (5 m/sec)
for a decade, then planets of Jovian mass and brown dwarfs will either be 
detected or ruled out for orbits with periods less than $\sim$15 years.

As specific examples, we outline the performance of our technique
on large amplitude and large eccentricity radial velocity signals
recently discussed in the literature and we delineate the region explored
by the measurements of 14 single stars made over a twelve year period
by Walker et al. (1995). Had any of these stars shown motion like that
caused by the exo-planets recently detected, it would have been easily 
detected. The data set interesting limits on the presence of brown 
dwarfs at orbital radii of 5--10 AU. The most significant features
in the Walker et al. data are apparent long term velocity trends in 
36 UMa and $\beta$ Vir, consistent with super planets of mass
of 2 $M_J$ in a 10 year period, or 20--30 $M_J$ in a 50 year period.
If the data are free of long term systematic errors, the probability of 
just one of the 14 stars showing this signal by chance is about 15\%.

Finally, we suggest an observing strategy for future large radial 
velocity surveys which, if implemented, will allow coverage of the
largest range of parameter space with the smallest amount of 
observing time per star. We suggest that about 10-15 measurements
be made of each star in the first two years of the survey, then
2--3 measurements per year thereafter, provided no (or slow) variation 
is observed. More frequent observations would of course be indicated
if such variations were present.

\end{abstract}


\section{Introduction}

For nearly two decades most high precision radial velocity surveys of 
nearby stars were focused on detecting radial velocity variations in
stars due to companions with mass and period of Jupiter. The signature 
would consist of changes in the relative stellar radial velocity with
a period of a decade and amplitude of a few tens of meters per second
or less, depending on orbital inclination with respect to the solar
system. The surprising recent result, triggered by the discovery
of 51~Peg~B by Mayor \& Queloz (1995), has been the finding that as 
many as 5-10\% of solar type stars have companions with mass 
$<10 M_J$ and with periods less than $\sim$3 years. No sub-stellar 
companions with periods longer than $\sim$3 years have so far been 
detected by radial velocity searches.

Are Jupiter mass companions at longer periods rare, or is it simply 
the case that current observations do not have the length or 
sensitivity to see them? Is the theoretical prediction by Boss (1995) 
correct, that Jovian planets should form preferentially at $>$4-5 AU 
separations from their primary? Our purpose in this paper is to show 
what we can learn from velocity data of a given duration and accuracy,
to help plan continued programs. 

The best measurement errors for a series of radial velocity measurements  
so far published are those of Butler et al. (1996), who observe a
magnitude $V=5$ star and quote an accuracy of 3 m/s for measurements taken
over one year. Measurements up until this work have been limited to
a lower accuracy standard of about 15 m/s.  Several other programs
(see section \ref{strategy} for a list of radial velocity search 
programs currently underway) are planning new or expanded searches with
a goal of obtaining measurements with similar accuracy. In light of
these efforts, and in expectation of their eventual success in obtaining
such accuracy, we shall use 5 m/s as a `canonical' value for the error
in many of the examples and the discussion below. Such advances in 
radial velocity calibration allow accuracy to be relatively free from
systematic error. Poissonian photon noise remains as the fundamental
limit to accuracy. In this limit, strong constraints can be placed
upon the existence of periodic radial velocity signals in a given 
set of data, given a suitable analysis technique.  

Many efforts have been made to determine whether a given set of data 
contains a signal. Most of those in common use are based upon the 
periodogram analysis techniques discussed by Scargle (1982). This 
technique is shown to be equivalent to a least squares fit for the 
signal at a given period, and he derives an exponential probability 
distribution of obtaining a false alarm from a given set of data.
Horne and Baliunas (1986 hence HB) have refined the technique by showing
that this exponential must be normalized to the total variance
of the data and derived an empirical expression for the number
of independent frequencies available to a set of data. Further
refinements (Irwin et al. 1989, Walker et al. 1995) account for
variable weighting of individual data points and correlations between
fitted parameters. 

Our work represents a different approach in which, rather than dealing 
with least squares minimization indirectly through a periodogram 
analysis, we examine the best fits to the data directly and determine
their significance.  We derive an analytic expression for the 
probability that a given best fit velocity amplitude is non-random.
We first develop analytical expressions relating sensitivity to 
planetary companions of different masses and periods, given velocity
measurements of specified accuracy, duration and number. Motion 
with periods longer than the duration of observations is detected 
with reduced sensitivity, and this reduction is explored by Monte 
Carlo methods. We illustrate our analysis technique by application 
to the published set of radial velocity data from Walker et~al. 
(1995), the longest time baseline survey so far published, with
quoted precision of 15 m/s. Limiting our analysis to the subset of
14 stars which have no known visual binary companion, we obtain 
quantitative upper limits to companions masses for orbital periods 
of a few days to periods as long as 100 years. Finally, we suggest
a strategy for efficiently implementing a search of a large number 
of stars for radial velocity signatures due to the presence of
a companion.

\section{Analysis Technique}

If a companion of mass $M_c$ exists in a circular orbit 
with inclination $i$ around its primary $M_*$, it will perturb the 
radial velocity of the star as observed from earth by: 
\begin{eqnarray} \label{vel-theo}
v(t) & = & {\left({{2\pi G}\over{PM_*^2}}\right)}^{1\over 3}M_c\sin{(i)}
                       \sin({{2\pi t}\over{P}} + \phi) \\
		      & = & K\sin({{2\pi t}\over{P}} + \phi),
\end{eqnarray}
where $K$ is the amplitude of velocity of the companion in a
circular orbit around its primary, $P$ is the period of the orbiting 
companion, $G$ is the gravitational constant and $\phi$ is an 
arbitrary phase factor. If an observer can detect the small temporal
changes in relative velocity due to a companion, then using fitting
or periodogram techniques, it becomes possible to derive a mass 
(or mass limit) for that companion.

Suppose that velocity data $v(t_i)$ have been obtained in observations 
extending over time interval $P_0$. For a given orbital period, $P$, 
we can perform a least squares fit to the data with the equation:
\begin{equation}\label{fit-eq}
v(t)= v_s\sin({{2\pi}\over{P}} t) + v_c\cos({{2\pi}\over{P}} t) + \gamma,
\end{equation}
to produce `best fit' values for the components of the motion $v_s$, 
$v_c$ and $\gamma$. At long periods and with a potential signal 
whose phase is unknown, the constant offset, $\gamma$, allows for
the possibility that a companion at a radial velocity extremum (ie.
near it's maximum or minimum) is properly modeled by the fit 
function.  For shorter periods ($P<P_0$) its inclusion or exclusion
has negligible effect so we will focus initially on this
domain.  Given fitted amplitude coefficients $v_s$ and $v_c$, a simple
trigonometric identity ($K = \sqrt{v_s^2 + v_c^2}$) 
yields the amplitude of the stellar velocity perturbation due to the 
companion.  From there we identify $K$ with the leading coefficient
in equation \ref{vel-theo} and invert to obtain a `best fit' companion 
mass:
\begin{equation}\label{mass-eq}
M_c = {{K}\over{\sin(i)}}\left({{PM_*^2}\over{2\pi G}}\right)^{1/3}.
\end{equation}
Fitting higher order harmonics would be used to refine the fit and
recover information about the orbital eccentricity of a companion.

The orbital inclination remains an unknown parameter in a set of radial
velocity data. Statistically speaking however, the average companion
mass of a set of systems randomly oriented in space which give amplitude
$K$ will be:
\begin{eqnarray}\label{mass-eq-ave}
\langle M_c\rangle={\int_0^{\pi/2} M_c\sin(i)di} 
     & = & \left({{PM_*^2}\over{2\pi G}}\right)^{1/3}
	      \int_0^{\pi/2}{{K}\over{\sin(i)}}\sin(i)di  \\
     & = & {{\pi}\over{2}}K\left({{PM_*^2}\over{2\pi G}}\right)^{1/3}.
\end{eqnarray}
Thus the average value for a companion mass is $\pi/2$ times
the directly derived $M_c\sin(i)$ value. Conversely, a companion with 
some mass will appear on average a factor of $2/\pi(\approx0.64)$ less
massive than its true value. Very large masses cannot be ruled out but 
do become increasingly improbable, with the probability that a given
mass, $M$, is exceeded being given by the formula:
\begin{equation}\label{m-exceed}
{\cal P}(M_c>M) = 1 - \sqrt{1 - \left({M_c\sin(i)}\over{M}\right)^2}
\end{equation}
For example, while values of the companion mass will be greater than twice
$M_c\sin(i)$ for 13\% of a large sample, the chance of a mass being greater 
than 10$M_c\sin(i)$ are only 0.5\%. The true companion mass will exceed
2/$\sqrt{3}(\approx 1.15)$ times the measured $M_c\sin(i)$ value in 50\%
of cases.

\subsection{Probability of a given velocity amplitude being 
exceeded by chance}\label{chance-prob}

In the absence of an unambiguous detection of a signal at some
period, we are faced with the question of whether a particularly large
fitted velocity amplitude at some period represents a real detection. 
Such spikes will occur, because the data are noisy, and the frequency 
analysis must be taken over a large number of possible periods (from a 
few days to many years). Adjacent fitted periods may have widely
different best fit velocity amplitudes even when the data have
no embedded signal.  What criterion can we apply to tell if a spectral
peak is improbably large compared to these noise spikes? More
generally, if the data for a star are analyzed in some way, what is the
probability that a given outcome would have occurred by chance? In this
section we obtain an analytical expression for the velocity amplitude 
(and hence companion mass) that will be exceeded by chance, with a
given probability and in a given frequency range.  

Suppose that in a given set of velocity measurements $v(t_i)$,
there is no real signal and that each measurement is drawn from a 
Gaussian distribution with mean zero and standard deviation $\sigma_p$.
The data can be fit to eqn. \ref{fit-eq} to produce coefficients 
of some amplitude ($v_s,v_c$). With no true signal, both $v_s$ and $v_c$
will be normally distributed about zero with standard deviation 
$\sigma_s=\sigma_c=\sigma$ and the phase of the fitted curve will be
uniformly distributed. For a set of $n_0$ measurements, taken randomly 
over a time period $P_0$, this assumption leads to the expression:
\begin{equation}\label{sigma-eq}
\sigma =  \sqrt{{2}\over{n_0}}\sigma_p.
\end{equation}
where factor $\sqrt{2/n_0}$ is derived from the least squares error 
analysis (see eg. Bevington and Robinson (1992) ch. 7) fitting a
periodic signal to random noise.

The probability of any data set with zero expectation value for $v_s$ and
$v_c$ to have any particular fit values is:
\begin{equation}
p(v_s,v_c) =  {{1}\over{2\pi\sigma_s\sigma_c}}
{e^{{-v_s^2}\over{2\sigma_s^2}}e^{{-v_c^2}\over{2\sigma_c^2}}}d{v_s}d{v_c}.
\end{equation}
which, converted to amplitude and phase gives:
\begin{equation}\label{prob-1}
p(K',\phi) = {{1}\over{2\pi\sigma^2}}{e^{{-K'^2}\over{2\sigma^2}}} 
		                          K' dK'd\phi.
\end{equation}
If we integrate this probability over all $\phi$ and from 
zero\footnote{For comparison, the integrated probability for a normal
random variable is given by:
\begin{displaymath}
{\cal P}= {{1}\over{\sqrt{2\pi}\sigma}}\int_{-K}^{K}
                        e^{{-K'^2}/{2\sigma^2}}dK'.
\end{displaymath}
} 
to some value $K$, we get the total probability, ${\cal P}$, of a fit with 
velocity amplitude $K$ or smaller:
\begin{equation}\label{conf-eq}
{\cal P} = 1 - e^{{-K^2}/{2\sigma^2}}.
\end{equation}
This probability applies to analysis of any single period. In
practice we are interested in the probability of a velocity amplitude
being exceeded by chance in a range of periods. If we assume that 
the probability of a given fit at one period is independent of 
every other period, then for $N$ periods the probability, $X$, that
no fit value exceeding a value $K_X$ will occur is the product of the
individual probabilities $X={\cal P}^N$, which to leading order gives
\begin{equation}\label{prob-eq}
X = \left(1 - e^{{-K_X^2}/{2\sigma^2}}\right)^N \approx
                         1 - Ne^{{-K_X^2}/{2\sigma^2}}.
\end{equation}
Higher order terms in the right hand equation converge to zero
as progressively higher power exponentials.  We can invert this 
equation to to derive a limit on the velocity as: 
\begin{equation}\label{v-eq}
K_X = \sqrt{-2\sigma^2\ln{\left({{1-X}\over{N}}\right)}}
\end{equation}
which expresses the velocity amplitude which will be exceeded
by any of $N$ fits to random data in a given period range
with probability $1-X$. 

The appropriate number of independent periods is related to the width 
of peaks in the frequency spectrum given by $df$=$1/P_0$. To be certain 
of sampling at a frequency that is close to the peak, we suppose 
that the sampling is made at frequency intervals $df=1/(2\pi P_0)$.
The number, $N$, of independent frequencies (or periods) in a given 
range is then given by:
\begin{equation}\label{n-eq}
N =  2\pi\left({{f_1 - f_2}\over{f_0}}\right)
\end{equation}
where $f_1$ and $f_2$ are the limiting frequencies of the range bounded
by periods $P_1$ and $P_2$ and $f_0=1/P_0$.

Finally, combining eqns \ref{sigma-eq}, \ref{v-eq} and \ref{n-eq}
we obtain an expression in terms of the accuracy $\sigma_p$, duration
$P_0$, the number of measurements $n_0$ and the probability $X$
that a velocity amplitude $K$ will be exceeded in a given frequency range:
\begin{equation}\label{vprob-eq}
K_X = 2\sigma_p\sqrt{{{1}\over{n_0}} 
        \ln{\left({{2\pi P_0\left(f_1 - f_2\right)}\over{1-X}}\right)}}.
\end{equation}
The value $K_X$ varies directly with $\sigma_p$ and varies with the 
inverse root of $n_0$, as we would expect from the central limit theorem.
Its sensitivity to the other parameters and our sampling assumptions 
depends on the details of the survey, but in general we will find that
the factor inside the natural logarithm is much greater than 1, so 
that a factor 2 change in any of the arguments produces only a small
fractional change in the value of $K_X$.

As an example, suppose that a high quality survey were made over a 
decade, with a total of $n_0=50$ observations per star and with 
rms accuracy, $\sigma_p=5$ m/s. For a false detection in a one
octave range around $P=4$ days, the velocity amplitude $K_X$ from 
eqn. \ref{vprob-eq} is 5.2 m/s. At 4 years, the amplitude is 3.9 m/s. 
If the star's mass is the same as the sun's, then from eqn. \ref{mass-eq}
we find these velocities for 1\% false detections will correspond 
to companions masses $M_c\sin(i)$ of 0.04 and 0.22 Jupiter masses
respectively. In practice, if a large number of stars are to be
sampled, say 100, and we would want a small probability of a false
detection in the sample, say 10\%, then we would want to decrease 
the probability to 10$^{-4}$ per octave per star. In this case, 
the mass limits increase to 0.047 and 0.28 $M_J$ for each range.
The small increase of only some $\sim$25\% is due to the fact 
that the argument of the natural logarithm in eqn. \ref{vprob-eq} 
is near 20 for this case.

\section{Monte Carlo Analysis}\label{monte-carlo} 

Eqn \ref{vprob-eq} will fail for periods longer than the span of
observations $P_0$, under conditions in which the data collection
is periodic, or if the total number of observations $n_0$ is too 
small.  This is because the windowing may imprint its own signature 
upon the derived best fit parameters and the assumption that random 
data are fitted with random phase breaks down. We devote this section
to a Monte Carlo analysis of synthetic radial velocity data, in order
to understand the regimes in which our analysis may fail and the 
manner of its failure. In this way we can eliminate false detections 
and establish the validity of a trend in the data consistent with a
true periodicity.

For our numerical experiments we assume radial velocity data are 
gathered for either 6 or 12 years. These data are spaced 
randomly in time subject to the constraints that data be `gathered'
during the same 6 month period of each year, that they be gathered 
only during 1/2 of each 29.5 day lunar cycle and that they be gathered 
only at `night'. We run a grid of nine Monte Carlo experiments 
varying the frequency of observation over 1, 5 and 20 observations per
year and the precision for each measurement over 5, 15 and 30 m/s.

We set velocities corresponding to the time of each observation using 
a Gaussian random noise term and the input error as:
\begin{equation}\label{vrand-eq}
v_{\rm sim}(t_i) = Rv_{\rm err}(t_i)
\end{equation}
with $R$ the random noise term and $v_{\rm err}$ is the error for each
point. The value for $v_{\rm err}$ is assigned as noted above.
We use the pseudo-random number generator `{\tt ran2}' provided by
Press et al. (1992) and the rejection method to create Gaussian random
numbers. In the analysis that follows, we fit a total of 3000 data sets
for the amplitude components $v_s$, $v_c$ and $\gamma$ for each star over
period ranges from 3 days to 100 years. The boundaries of each period 
range are defined in table \ref{range-tab}.  We increase each 
successive fitted period by an amount such that the total number of 
orbital cycles over the full observation length decreases by 1/2$\pi$ 
(1 radian) as in the analysis above or the period increases by 
1/5~year, whichever gives the smaller interval. The chance of any 
particular outcome is given by the fraction of the synthetic data 
sets with that outcome.  

\subsection{Confirmation of the Analytical Results}

For the subset of experiments with assumed 5 m/s precision, figure 
\ref{mc-anal-cmp} shows the radial velocity amplitude for each fitted period 
which is exceeded in 1\% of the Monte Carlo trials, ie. there is a 99\% 
probability that a specific period analyzed will not exceed this value.
Experiments with higher or lower assumed precision produce limits
scaled upward or downward on the plot but otherwise show the same
qualitative features.  Also included are the $N=1$ limits provided 
directly by eqn. \ref{v-eq}.  In general, the Monte Carlo results 
confirm the validity of the analytical results above. The difference
between $N=1$ analytic and Monte Carlo results varies about 2\%, 
consistent with statistical fluctuations, except at the assumed windowing
periods and at periods longer than $P_0$. The analytical prediction for
the experiment with the most sparsely taken data (1 measurement per 
year for 12 years) lies some $\sim15\%$ below the Monte Carlo result
for periods less than 2 years, but agrees to $\sim2\%$ over the 
remainder of the valid period regime.  

A comparison of the limits provided by the analytical (eqn.
\ref{vprob-eq}) and Monte Carlo methods for each period range noted
in Table \ref{range-tab} are also shown in figure \ref{mc-anal-cmp}.
The assumed windowing periods are masked out of each of the Monte Carlo
limits and the results represent limits based on the remaining portion
of each range. In general, the Monte Carlo results again confirm the
validity of the analytical result to within a few percent, with the
exception of the series with only one measurement per year. In that
case, the Monte Carlo experiment produces limits which are some 50\%
or more larger than eqn. \ref{vprob-eq} predicts.  

We consider in turn in the sections below the differences between
the analytical derivation and the Monte Carlo results due to the long
period fall off in sensitivity, due to small numbers of observations,
and due to the inherent windowing in the data. 

\subsection{Loss of Sensitivity at Long Periods}\label{long-per}

The results shown in figure \ref{mc-anal-cmp} show that at periods 
longer than the 12 year window the sensitivity to a velocity signal 
drops off in very nearly power law form. In light of this
behavior, we adapt an {\it ad hoc} prescription for the velocity limit
using the eqn. \ref{vprob-eq} result at short periods and a power law
at longer periods as
\begin{equation}\label{vprob-eq-long}
\hat K_X = \cases{ K_X                  & for $P<\beta P_0$;\cr
                   K_X \left({{P}\over{\beta P_0}}\right)^\alpha
                                        & for $P>\beta P_0$; \cr }.
\end{equation}
We then fit for the free parameters $\alpha$ and $\beta$ and thereby
recover limits for periods much longer than that of the observing
window. In this equation, we assume that the value of $K_X$ used
for long periods ($P>\beta P_0$) is that defined by the last period
range prior to the onset of the fall off. This assumption ensures a
smooth joining of the two regimes.

\begin{figure}
\plotone{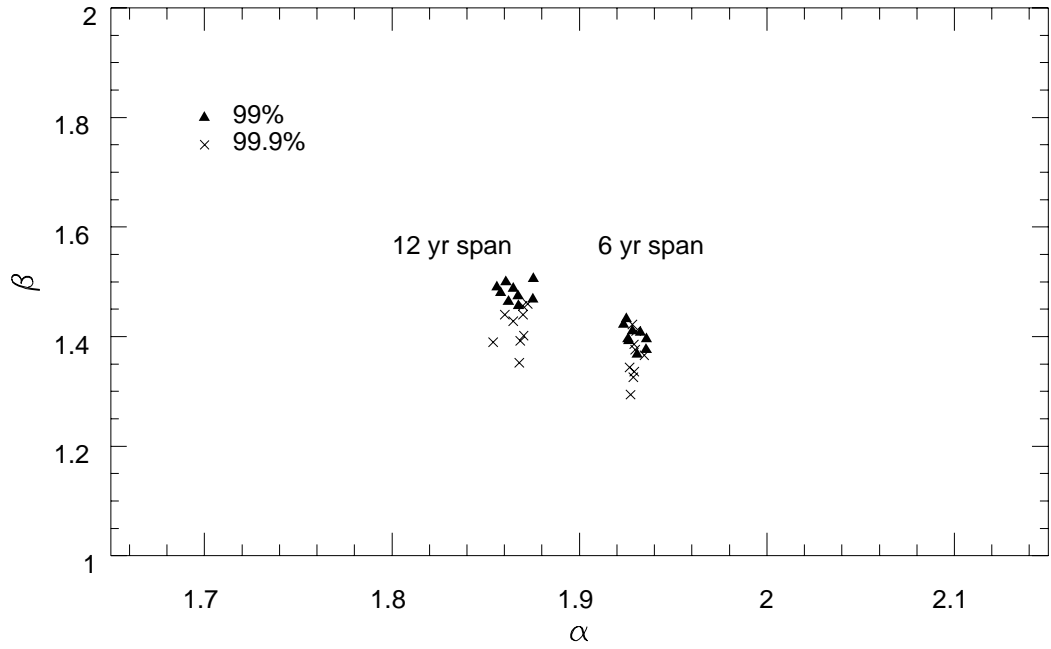}
\caption{\label{mcl-bestf}
Best fit values for the long period sensitivity fall
off parameters $\alpha$ and $\beta$ for the nine monte carlo
experiments run for both 6 and 12 year baselines.}
\end{figure}

We fit the Monte Carlo results for the constants $\alpha$ and 
$\beta$ in eqn. \ref{vprob-eq-long} for each of the experiments and plot
their values in figure \ref{mcl-bestf} for both the six and twelve year
observing windows studied. The fitted values for the 12 year window are 
typically:
$$
\alpha \approx 1.86
$$
with the turn off in sensitivity beginning between 
$$
1.4 \leq \beta \leq 1.45
$$
for the 99\% probability curve and similar values for the 99.9\%
probability curve.  A slightly steeper power law exponent 
($\alpha\approx 1.92$) is found for a 6 year window. 
If we err on the side of caution and assume that the turn-off occurs at
the {\it small} end of the range (by setting $\beta=1.3$, for both the
99\% and 99.9\% probabilities), then we provide slightly more conservative 
limits than the best possible based on our Monte Carlo analysis. Under this 
assumption, we have included in figure \ref{mc-anal-cmp} the long period 
fits for the velocity limits placed upon the data by eqn. \ref{vprob-eq-long},
and the shorter period limits for 11 period ranges less than $1.3\times$12 
years. 

\subsection{Limits of Sparse Data}\label{sparse}

When a data set contains only a few measurements, a least squares 
analysis will depend strongly upon the measured value and placement 
in time of each measurement. How many data are needed to assure that
the random data/random phase assumption is reliable and we are able 
reproduce the results of equations \ref{v-eq} (with $N$ set to unity)
or \ref{vprob-eq} (for octave period ranges)? Is there a difference in 
the number of measurements that must be made if we assume a strategy 
of taking, say, one or two measurements per year over a long period 
or taking several measurements per year but over a much shorter
baseline?

Taking the first strategy, we assume the data are gathered over a 6
year span with an error in each measurement of 5 m/s. If a star is
observed with a frequency of one observation per year, we find 
(figure \ref{vlim-sparse6}) that the eqn. \ref{v-eq} limits with
$N=1$ underestimate the Monte Carlo results by more than a factor of 
two for periods shorter than 1 year, and by a smaller margin at all 
periods. The same experiment with a 12 year span shown in figure 
\ref{mc-anal-cmp} shows a much smaller ($\sim$15\%) difference. 
Increasing to three observations per year for 6 years the analytic
equation underestimates the limits by $\sim$10\%, while 5 measurements 
per year duplicates the analytic results to 5\% or better.

For octave sized ranges, the analytical and Monte Carlo results converge
somewhat more slowly. Figure \ref{mc-anal-cmp} shows that a single
measurement per year over 12 years is sufficient only to provide limits
a factor of two higher than would be predicted analytically for periods 
less than 2 years. When data are gathered at the higher rates shown 
(5 and 20 obs/yr), the agreement is excellent. An experiment with 
two measurements per year (not shown), for a total of 24 measurements, 
is sufficient to recover the analytical form to $\sim$10\% in all 
period bins. With the six year baseline shown in figure 
\ref{vlim-sparse6}, agreement at the $\sim$20\% level is reached if
three measurements per year (18 total) are taken.

\begin{figure}
\plotfiddle{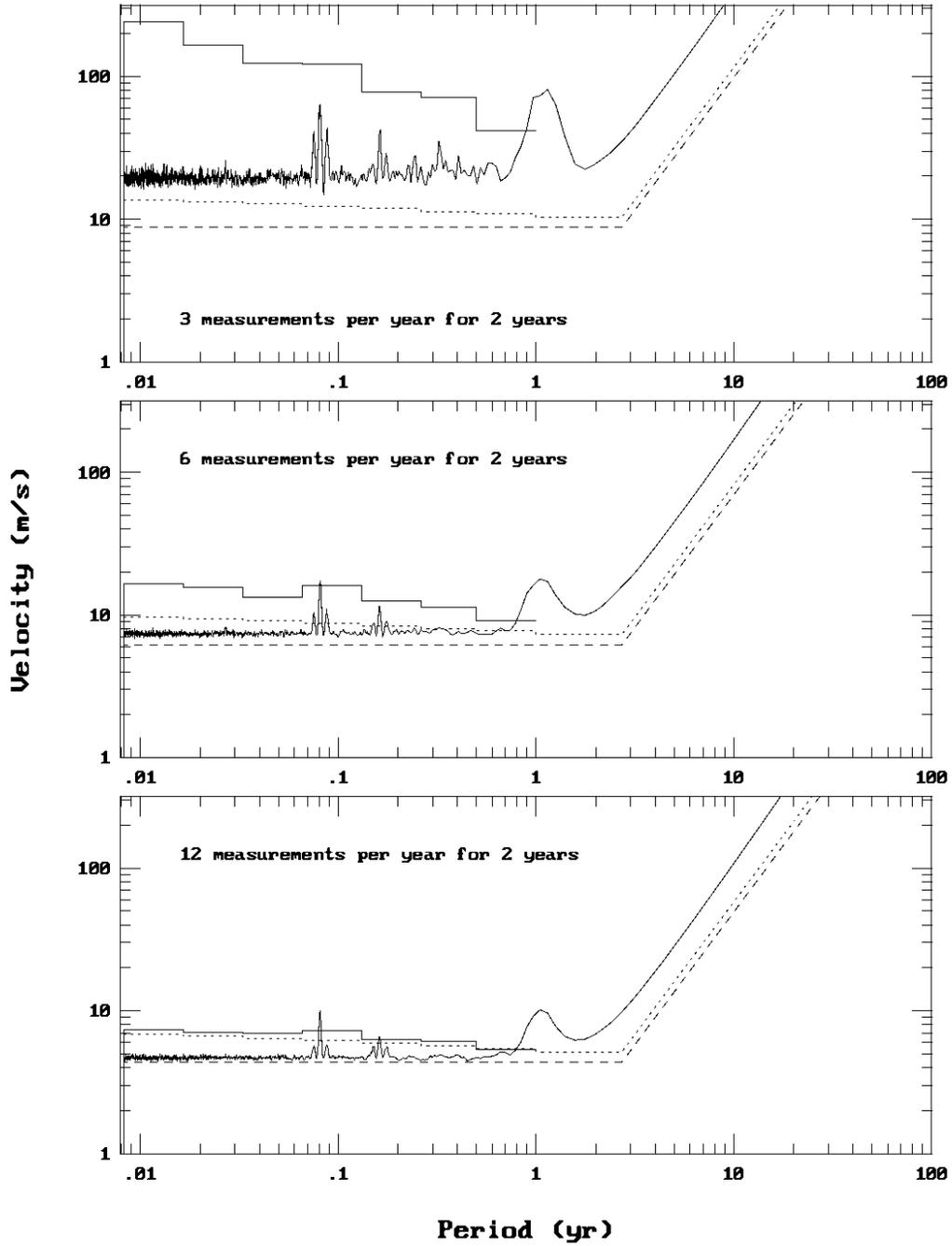}{6.79in}{0}{72}{72}{-230}{-20}
\caption{\label{vlim-sparse2}
Velocity limits given by monte carlo simulations (solid), 
single period ($N$=1) analytic limits (dashed). A 2 year window is
assumed with 3 (top), 6 (middle) and 12 (lower) observations per 
year. As before, histograms represent the monte carlo (solid) and
analytic (dotted) limits for octave sized ranges and each of the
histogram limits ignore the periods affected by the assumed lunar
and annual windowing of the data. In this plot however, for periods 
longer than 1 year the monte carlo histogram limits are suppressed.}
\end{figure}

Taking the second strategy, we assume data are gathered over a
two year window. We do not believe we can rely upon octave range
limits for such a short data gathering period because of the large 
effects of windowing, which we discuss below. The long period fall
off is similarly affected. We therefore limit our discussion for these
experiments to limits for individual periods, shorter than about one 
year. With a two year data window and a total of 6 measurements (three
measurements per year), we again find (figure \ref{vlim-sparse2}) that 
the Monte Carlo limits exceed those of eqn. \ref{v-eq} with $N=1$ by 
more that a factor of two.  Increasing to 6 measurements per year (12 
total), we lose only 15\% of the maximum sensitivity for $N=1$, while 
12 measurements per year (24 total) recovers the analytic results with 
only a $\sim$5-7\% difference.

In order to obtain limits which retain the benefits of a given 
precision to within 15\% at any single period, we find that at
least $\sim$12 or more observations of a star must be made. This 
number of observations produces limits a factor of two larger than 
predicted over octave ranges.  To reduce the difference
to $\sim$5\% for a single period and 15\% over octave ranges
requires at least 18-20 measurements. Barring windowing effects, 
these minimum requirements do not seem to depend strongly upon the
time span over which the data were gathered, but only upon their 
accuracy and number.

\subsection{Windowing}\label{windowing} 

Sensitivity loss of a factor of two or more is present in `blind spots' 
for any single period near the assumed lunar and annual windowing periods
for every experiment performed. There are also double period counterparts
and beat periods between the lunar and annual data windows, though lower
sensitivity loss is evident there. Day/night windowing effects are not 
visible in the limits due to their extreme short periodicities. When the 
data are sparse and the data are gathered over a short period $P_0$, the 
effects are especially pronounced.  Figure \ref{vlim-sparse2}) shows
that for a period $P_0$ of two years, the lunar windowing effects are
observable not only at the lunar orbital period, but also at the double, 
triple and quadruple period aliases. Additionally, fitting for the long
period turn-off becomes of little use because the turn-off occurs at 
a period with lower sensitivity than can be modeled analytically. 

Based on these results, we suggest that the limits which can be placed 
on signals at periods corresponding to a lunar or annual windowing period 
cannot be reduced beyond a factor of two greater than that given by eqn.
\ref{v-eq} with $N$ set to unity at a windowing period or a
factor of $\sim$3/2 at one of its double or beat period counterparts.

\section{Comparison to Periodogram Techniques}\label{compare}

To obtain definitive probability that a signal that been detected at
some period is nonrandom, nothing less than a full Monte Carlo
analysis is adequate. For a large survey which is continually updated
as more data are gathered, such analysis is unfeasible because
of the considerable commitment of computational facilities to perform
a statistically meaningful analysis. Even for the computers of today,
a sample of 500 stars might prove unmanageably burdensome. To reduce
the effort required per star, either periodogram or fitting techniques
such as ours may offer a lower cost alternative.  We will now make a
comparison of our technique to periodogram techniques in common 
use.

Each technique is based upon a $\chi^2$ analysis of the data.
Indeed, for equally weighted data least squares analysis and 
periodogram analysis have been shown (Scargle 1982) to be equivalent.
The main difference lies in the fact that on the one hand,
a periodogram utilizes a normalized measure of the power of the signal
at some period while our technique relies directly on the value of 
the best fit velocity amplitude. Additionally, with the present 
analysis, we allow the data to be fit with unequal weights, though 
the amplitude limits derived are based upon only upon equally 
weighted data.

Let us examine the least squares fitting procedure and, for 
purposes of illustration, limit ourselves to the case of fitting 
for only the coefficients $v_s$ and $v_c$ in eqn. \ref{fit-eq}.
In this case, the best fit coefficients derived from the $\chi^2$
minimization at some frequency $\omega=2\pi/P$ for a set of 
$n_0$ velocity measurements, $v_i$, are: 
\begin{equation}
v_s =  C_{ss}\sum_{i=1}^{n_0}{ {v(t_i)\sin \omega t_i }
                     \over{\sigma_i^2}}
           + C_{sc}\sum_{i=1}^{n_0}{{v(t_i)\cos \omega t_i }
                     \over{\sigma_i^2}}
\end{equation}
and
\begin{equation}
v_c =  C_{cs}\sum_{i=1}^{n_0}{{v(t_i)\sin \omega t_i }
                             \over{\sigma_i^2}}
           + C_{cc}\sum_{i=1}^{n_0}{{v(t_i)\cos \omega t_i }
                             \over{\sigma_i^2}}
\end{equation}
where the subscripted $C$ terms are the four components of the 
covariance matrix used to derive the fit (see for example Press et al. 
1992 ch. 15.4 for a discussion). When these terms are combined to
form the velocity amplitude $K$ as $K=\sqrt{v_s^2 + v_c^2}$ and data
are translated in phase by a value $\tau=\tan^{-1}(v_c/v_s)$ (derived 
by setting $C_{cs}=C_{sc}=0$) then, as was shown by Lomb (1976),
the square of the best fit velocity amplitude, $K^2$, becomes the 
unnormalized power of the periodogram at that period. With the
identification of $K^2$ with the periodogram power, we note that 
false alarm probabilities are given in each case is given as an 
exponential of $K^2$, with a normalization given by the variance,
$\sigma^2$, of the data.

The use of the velocity amplitude rather than a normalized measure
of its square represents an improvement to existing techniques for 
several reasons. First, a physically meaningful limiting velocity amplitude
(or equivalently, a companion mass $\times\sin(i)$) is explicitly a part 
of the definition of the probability. A potential weakness of this method 
is that because it utilizes amplitude as a figure of merit rather than 
power, its dynamic range is more compressed on a given plot. A single 
dominant peak will not stand out to nearly the extent that occurs in a 
periodogram. In spite of this somewhat minor defect, we submit that
a best fit amplitude is a far more useful quantity to an observer than 
is the power.

In sections \ref{sparse}-\ref{windowing} we have outlined the
regimes for which our analysis is valid and the manner in which it
fails for sparse or windowed data and for very long periods. Because 
of the similar origins of our analysis and periodograms we expect that 
similar failure modes also apply to periodograms. Hence probabilities
derived from sparse data ($n_0\lesssim 10-15$) and at `windowed'
periods such as the annual cycle using a periodogram will yield erroneous
results. Extensions to standard periodgram techniques (Irwin et al. 1989
and Walker et al. 1995) which explicitly account for unequal statistical
weights and correlations between fit parameters may provide more accurate
limits than our eqn. \ref{vprob-eq} in such regimes. 

Our extensions to long periods explicitly provide limits on
the amplitude of the signal (and therefore $M\sin(i)$) possible at
any given period at least 10 times as long as the data window. The 
limits account for the fact that a long period signal may in fact be
near an extremum during the time over which most or all of the data 
were gathered.

Both techniques may be used to determine the probability of
a signal being nonrandom for a single period, for a period range
or over all independent periods. The Scargle (1982) and HB false 
alarm probability generates the probabilities, in the ideal case, by
requiring a Monte Carlo analysis to specify the number of independent 
periods, $N$. Their analysis to determine $N$ is limited to sampling 
frequencies below the Nyquist limit however. With unevenly sampled data, 
it is well known that higher frequencies are accessible without aliasing. 
How far above the Nyquist limit a signal can be detected and how many
additional independent frequencies (if any) are required remains 
unknown. 

We also require a specification of the number of independent 
periods, however our analysis uses a definition of the number
of independent periods (not equivalent to the HB definition) based
only upon the width of a spectral peak. We make no distinction between
potentially aliased spectral peaks at high frequencies and those
found at lower frequencies. The excellent correspondence between
our analytical formalism and our Monte Carlo analysis for each period
range shows that the definition of $N$ made in eqn. \ref{n-eq} is 
reasonable. The functional dependence of the amplitude limit on $N$ is 
quite weak, going only as $\sqrt{\ln{N}}$. When $N$ is large, as is the 
case for the shortest period bins, our definition will yield slightly
more conservative (higher) limits than the comparable HB limits, while for 
longer periods when $N\lesssim 100$, our limits may be somewhat lower. 

\section{Application to Real Data}\label{realdat}

In this section we apply our analysis technique to data for 
two stars obtained by Mayor and his collaborators at the Geneva 
Observatory, data obtained by Marcy et al. (1997) for the star 51 
Pegasi, and to the data obtained by Walker et al. (1995) in their 
12 year search for extra-solar planets. The Walker et al. radial
velocity data are for a set of 21 stars with data taken over a 12 
year period from 1980-1992. The data used in our analysis were 
originally archived at the Astronomical Data Center 
(URL http://hypatia.gsfc.nasa.gov/adc.html) by Walker et al. upon
publication of their work.  We limit our analysis to the subset of 
14 stars for which no visual binary companion is known, shown in 
Table \ref{star-tab}.  For these stars, no other periodic radial velocity
signatures which could obscure a planetary signature are present, and 
no significant periodicities attributable to planetary companions 
were found by the Walker et al. search. 

Using equations \ref{mass-eq} and \ref{vprob-eq} we can derive for 
any period (or period range) of interest the limit below which random
data is fit with probability $X$ to be: 
\begin{equation}\label{m-eq}
M_c = {{{\hat K_X}\over{\sin(i)}}
	      \left({{PM_*^2}\over{2\pi G}}\right)^{1/3}} 
\end{equation}
where we assume the orbital period, $P$, is at the midpoint of some
range of periods shorter than $\beta{P_0}$ or that $P>\beta{P_0}$.
This mass limit depends upon both the velocity amplitude limit $K_X$, which 
changes slowly, and also the period for which we fit the data. The minimum 
mass detectable by a set of measurements increases only as the cube root 
of the period at short periods, but for $P>\beta{P_0}$, this dependence 
becomes much steeper, increasing faster than $P^2$.

Once we have mass or velocity amplitude limits for a set of data,
we can define quantities $M_{99}$ and $M_{999}$ via eqn. \ref{m-eq} as 
the mass exceeded by chance by a fit at the 1\% and .1\% level of 
probability in each period range. For a sample of radial velocity 
measurements of say 10 stars observed for 10 years divided up into 10
period ranges, these values are of interest because if there are no 
true periodicities in the data, we would expect to find by chance one
apparent planet with best fit mass $M > M_{99}$, but to find a mass
greater than $M_{999}$ with only $\sim$10\% probability.

\subsection{Determining the Measurement Uncertainty}

In order to obtain a value of $\sigma_p$ for use in eqn. \ref{vprob-eq}
or \ref{m-eq}, we assume that no strong periodic signals are present 
in the data and that each datum is drawn from the same statistical 
distribution. Then we may use the rms scatter of all the data for 
a star as an empirical measure of the error, $\sigma_p$, for each
measurement of that star. For data in which no clear signal is 
observable, this measurement of the error will give a more reliable
estimate of the true value than from internal estimates.  In Table
\ref{star-tab}, we show both $\sigma_p$ as derived directly from the
data as well as the average of the internal errors for each star 
($\sigma_i$) quoted by Walker et al.

\subsection{Detecting Large Amplitude and Large Eccentricity Signals}

In this section we show that our analytic technique is capable of
detecting large amplitude signals and signals with high eccentricity.
As an initial test we obtained the data for the original discovery 
of the companion to 51 Pegasi, taken by Mayor and his collaborators
at the Geneva Observatory. These data consist of the original 35 
measurements as published by Mayor and Queloz (1995) as well as
their observations of the star since that time. The total number
of observations used in our analysis was 89 radial velocity 
measurements made over 2.4 years with internal errors of 15 m/s.
The value for $\sigma_p$ was obtained from the rms scatter of all
of the velocity measurements and was $\sigma_p = 44.5$ m/s. An 
independent set of measurements (Marcy et al. 1997) was also used
to compare the technique using higher precision data.  A total of 116
measurements are characterized by an rms scatter of $\sigma_p$=40.6~m/s
and were gathered over a total time span of 325 days with internal 
errors of 5 m/s.

\begin{figure}
\plotfiddle{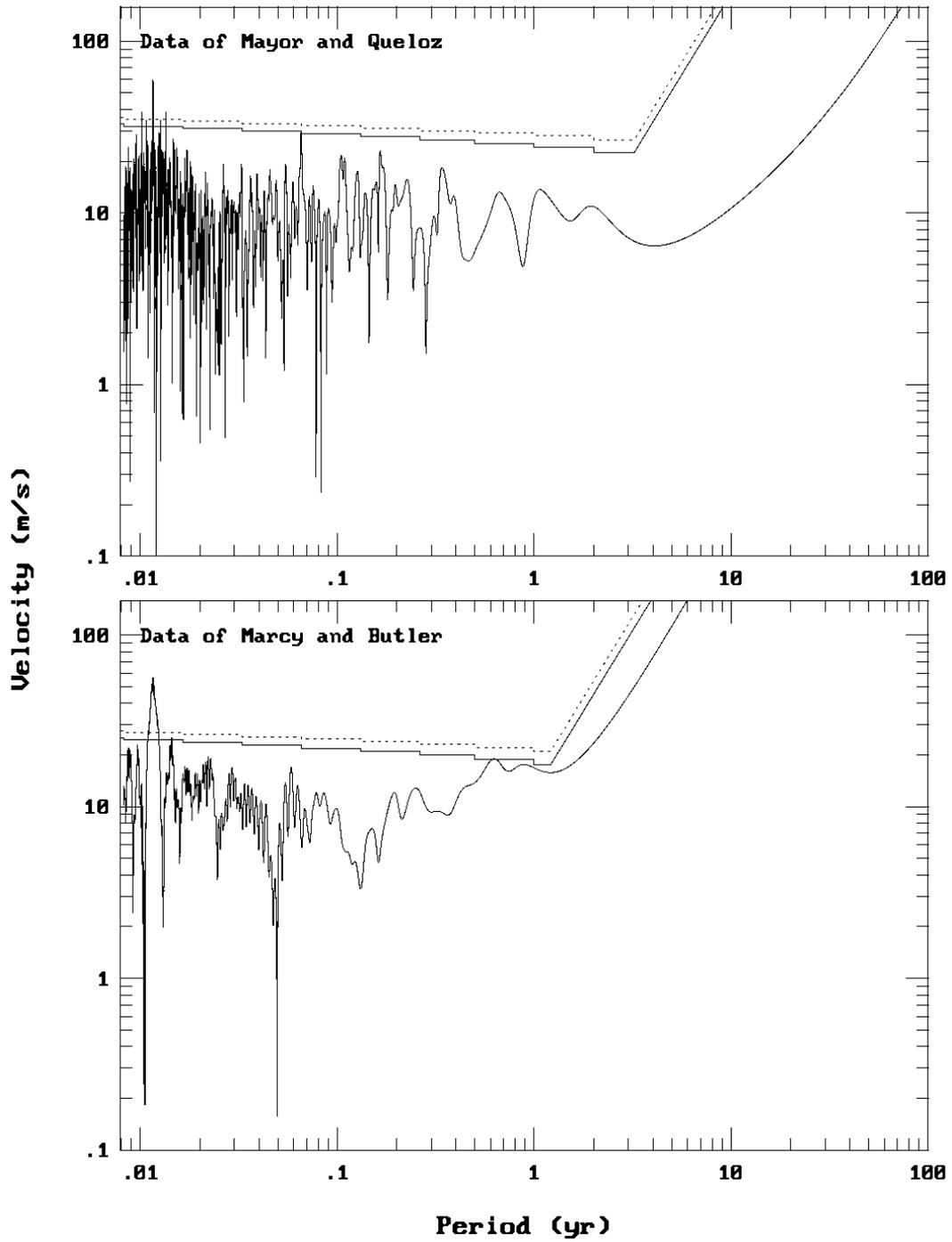}{6.79in}{0}{75}{75}{-230}{-20}
\caption{\label{51peg-vel}
Best fit velocities for the star 51 Pegasi. The solid and dotted
histograms denote respectively, the limits below which random data 
would occur with 99 and 99.9\% probability. }
\end{figure}

We show the results of these two tests in figure \ref{51peg-vel}.  
For each set of measurements, we detect a clear peak in the best 
fit velocity amplitude at a period $P=4.23$ days, as expected. 
For the Mayor and Queloz data, we also detect a number of side
lobes peaks which represent the radial velocity signal `beating'
against other periodicities in the data such as the 29.5 day 
lunar cycle (the data were taken predominantly during the same 
half of each lunar cycle). The Marcy et~al. data show one 99\%
significant peak just shortward of one year. We consider this
peak to be an artifact of the short time baseline of their data
(less than one year) and do not consider it very significant. 

A second 99\% significant period is detected at 23.84 days in the
Mayor and Queloz data. We have re-analyzed the residuals  of the
measurements (with the 4.23 day periodicity removed) and found
that the peak remains and so cannot be attributed to an alias
of 4.23 day periodicity.  Is it an artifact of the rotational
period of the star itself? We note that the period is roughly
in a 2:3 ratio to the observed 37 day rotation period for 51 Peg 
(Henry et al. 1996). We suspect that with the incomplete phase 
coverage for periods near 24 and 37 days, the stellar rotation 
period may be aliased to the observed 23.84 day period. Only 
complete phase coverage may be able to determine the origin of 
this signal.

The independent radial velocity observations of Marcy et al.
do not show a similar periodicity and the precision of their
measurements is only half the best fit amplitude from the Geneva 
data. In their work more than 3/4 of the data were gathered in
less than two rotation periods of the star. In such a case, it 
is unclear whether a rotation signature would be observable in
Doppler spectroscopy data. 

As a second test, we obtained another set of radial velocity data 
from Mayor. In this case the data were obtained under the condition
that the identity of the data and whether they contained a signal
not be disclosed until the conclusion of the test. These data consisted 
of 45 measurements taken with the CORAVEL spectrometer with internal 
errors of 300 m/s. The data were gathered over $\sim$15 years and 
the value for $\sigma_p$ obtained from the rms scatter in the velocity
measurements and was $\sigma_p = 800$ m/s.

\begin{figure}
\plotone{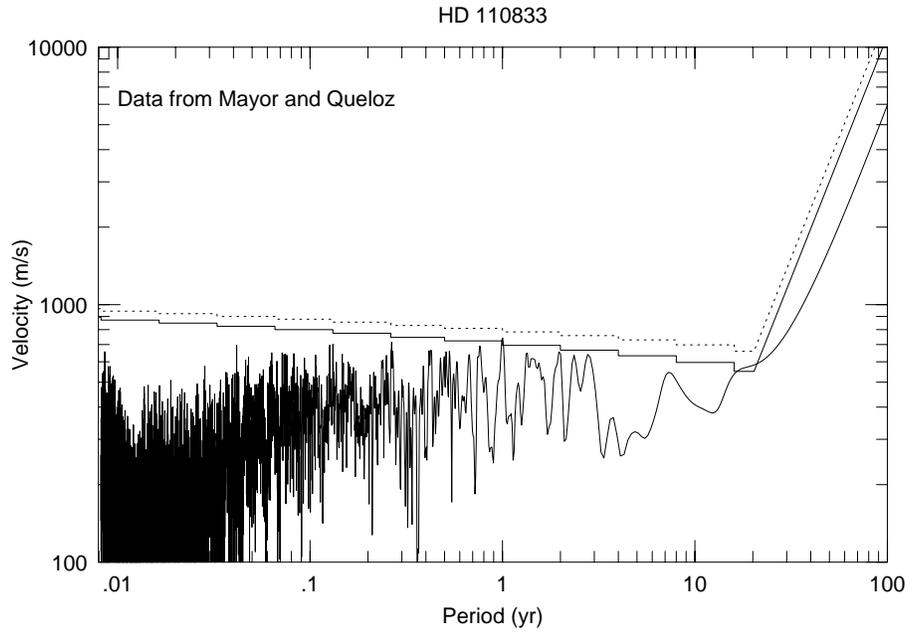}
\caption{\label{mayanon}
Best fit velocities for the data derived from star HD 110833.
A 15 year periodicity is apparent in the best fits with $>$99\%
probability that it is nonrandom.}
\end{figure}

The best fit velocities and the corresponding 99 and 99.9\%
probability limits for the fits are shown in figure \ref{mayanon}.
In this case, we were unable to detect a significant periodicity
in the data except a possible long term signal near 15 yr.  The data
show no obvious periodicity in the velocities for orbits of $\sim$15 yr,
but do show that several measurements are some 3--5 $\sigma$ away from 
the mean of any other velocity measurements of the star. These data
were gathered within 6 days of each other in the fall of 1996.

\begin{figure}
\plotone{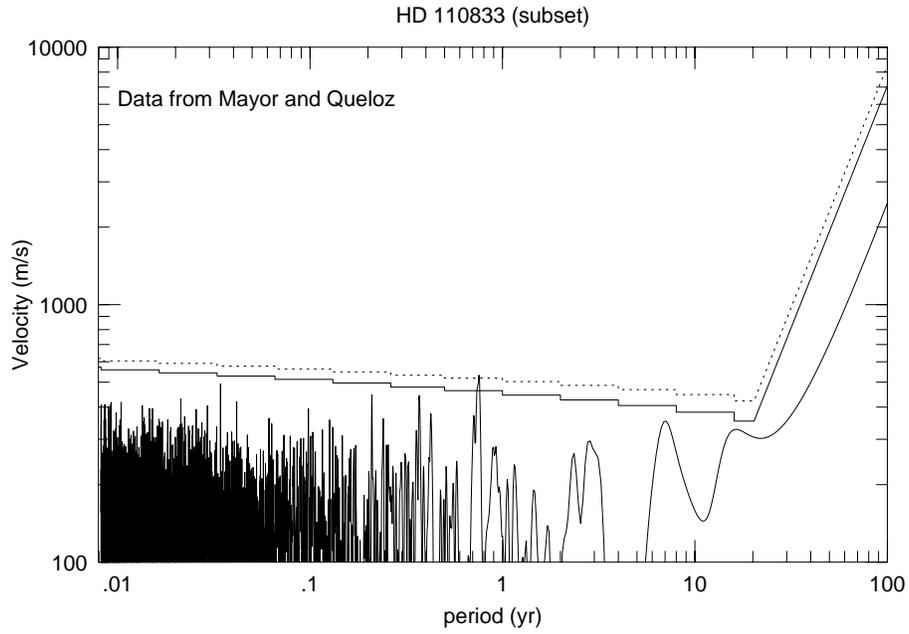}
\caption{\label{mayansub}
Best fit velocities for the the subset of the data from HD~110833 
with 5 measurements removed. In this case, a periodicity is
present at 275 days with probabability $>$99.9\% that it is 
non-random. }
\end{figure}

To test the effect of these data we deleted them from the sample and
reapplied our analysis. In this case, the rms scatter was reduced 
to $\sigma_p = 480 $ m/s and $n_0$ to 40 while the duration of the
measurements $P_0$ remained the same. Figure \ref{mayansub} shows
the results of this reanalysis.  In this case a peak in the best
fits for a circular orbit, well above the 99.9\% probability curve,
becomes visible at 275 days.  With the detection of this peak, we
concluded our test and obtained the identity of the star from which
the observations were taken. The star from which the data were 
obtained was HD~110833, for which Mayor et al. (1996) published an
orbital solution with a best fit companion mass $M\sin(i)=17~M_J$,
a period of $P=270$ days and an orbital eccentricity $e=0.7$ using
data from both the CORAVEL and ELODIE spectrometers. The difference
in the period derived from our analysis and the Mayor et~al. fit 
is due to the inconsistency between the high eccentricity of the 
companion and assumption made in our analysis that the orbit is 
circular.

In discussions with M. Mayor and D. Queloz, several issues were brought 
forward. First, the data from the fall 1996 run may have been faulty due to 
a combination of several factors, including a slight misfocusing of the 
telescope or a temperature instability in the spectrometer itself. However,
while the measurements in question are unusually distant from the mean of 
the other measurements for that star, they were gathered during a single 
periastron passage of the companion and therefore may not contribute as 
strongly to the detection of the periodicity as they would otherwise. A 
fit for the set of orbital elements would then yield a higher eccentricity 
than is truly the case. 

This case may therefore expose a degeneracy between results using data derived 
from a companion which is truly in a highly eccentric orbit and data 
with possible systematic biases. Our technique is based upon only the lowest 
order Fourier components of the signal (i.e. a circular orbit) and does not 
account for eccentric motion.  The fact that so many data (5 of 45 measurements) 
were so far from the mean and that they were from the same observing run 
suggests that removing the data from that run from our analysis is 
justified. Omitting them, we are able to recover a strong periodicity near
the best fit period for the orbital solution.

With the results from this section we can be confident that our analytic
technique is capable of detecting signals with large amplitude and/or
large eccentricity.

\subsection{Masses from Best Fit Velocities of the Walker et al. 
Sample, and Analysis of Significance}

\begin{figure}
\epsscale{0.8}
\plotfiddle{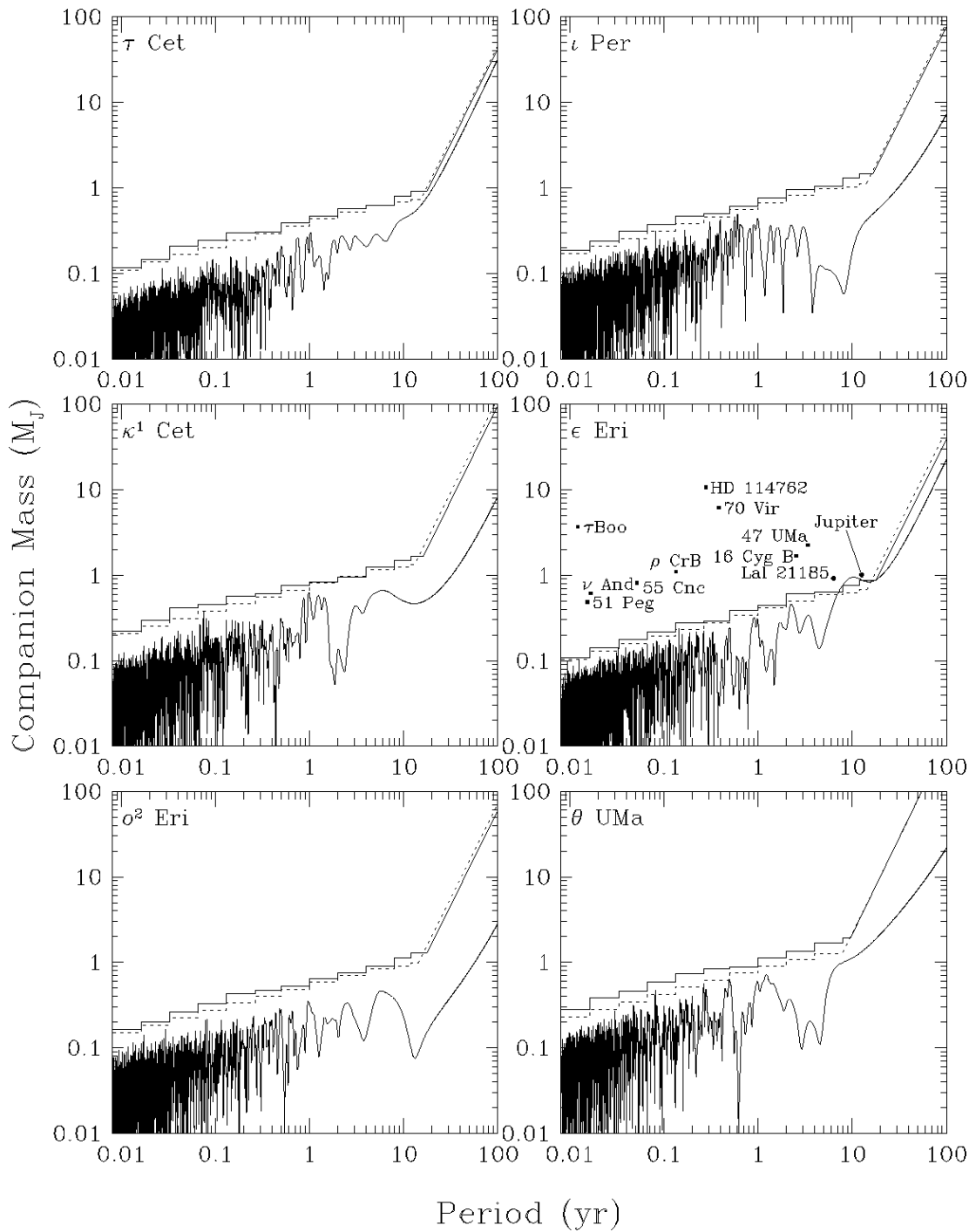}{6.95in}{0}{73}{73}{-230}{-20}
\caption{\label{datamc-bf}
Best fit companion masses ($\times \sin(i)$) for each star over a 
range of periods corresponding to the most sensitive range of the 
data. The histograms represent the 99\% mass limits for each
specified period range as given by our analytic formulation (dotted 
histogram) and based on a monte carlo experiment (solid histogram) 
consisting of 3000 simulated data sets.  Also plotted (squares) 
are the measured $M_c\sin(i)$ values of recent planet detections 
(see text). Jupiter and the astrometrically detected (but unconfirmed)
companion to Lalande 21185, are shown omitting a $\sin(i)$ 
correction (circles).}
\end{figure}

We have shown that our analytic technique is capable of detecting 
large amplitude periodicities in radial velocity data. In this
section we move to lower amplitude signals and upper limits to
companion signatures. For each star in the Walker et al. sample,
best fit velocities for periods between 3 days and 100 years were 
determined by the least squares method using eqn. \ref{fit-eq}. 
Statistical weights for each datum were taken to be 1/$\sigma_i^2$,
where $\sigma_i$ is the quoted internal error for each point
s as given by Walker et al. Plotted in figure \ref{datamc-bf} are the
corresponding companion masses $M_c\sin(i)$ obtained from eqn. 
\ref{mass-eq} using the stellar masses from table \ref{star-tab}. In 
order to determine the significance of any particular best fit value 
we compare the best fit mass for some period to the limit provided by 
our analytical analysis and to Monte Carlo experiments similar 
to those described in section \ref{monte-carlo} for synthetic
data. Each of these limits are shown in figure \ref{datamc-bf}.

While as expected from the results of the original analysis of 
Walker et al., there are no clear cut companion signatures, in 
several cases the data produce statistically significant fits. We 
tabulate each of these periods in table \ref{signif-tab}. Are any 
of these signatures due to the existence of a companion? Stellar 
processes such as pulsation, rotation or magnetic cycles can affect
the measured radial velocity for a star and in many cases it is 
quite possible to fit such signals with orbital solutions. Early 
in this century for example (see eg. Jacobsen 1925, 1929), the 
radial velocity variations of Cepheid variable stars were fit 
with Keplerian orbits. Although today no one would attribute
Cepheid radial velocity variations to a companion, the principle
that processes intrinsic to the star must be eliminated from 
consideration remains if we are to be certain that a given 
radial velocity detection is definitely due to a companion.

Many of the signals in table \ref{signif-tab} do in fact correlate
with known periodicities due to stellar rotation or magnetic cycles
in the star. For example, we find a $>$99.9\% significant $\sim$10 yr 
period in $\epsilon$ Eri and two short period signals (at 11.9 and
52.5 days) with $>$99\% probability. Walker et al. establish that the
10 year and 52 day peaks are aliases of each other and McMillan et~al.
1996 have definitely connected this periodicity to a stellar magnetic
cycle.  Gray and Baliunas (1995) have observed an 11.1 day periodicity 
in the Ca H\&K S-index with an extensive data set. They comment that
subsets of their data taken during different observing seasons produce 
peaks varying in period from 11 to 20 days. We conclude that we
are seeing a comparable effect in the Walker et al. radial velocity 
data and are in fact detecting the rotational signature of the star.  

Further work by the same group (Gray et al. 1996) on the star 
$\beta$~Com provides evidence of a magnetic cycle. However their 
measurements have sufficient duration only to have observed a minimum, 
and a period is not known. Figure \ref{datamc-bf} shows that for 
$\beta$~Com the probability that the best fit velocity amplitude
is not random exceeds 99\% for periods near 10 years. Assuming
a 10 year period, we calculate that the best fit radial velocity 
curve went through its minimum in 1988/1989, which is coincident with 
the photometric and Ca H\&K minimum observed by Gray et al. (1996).
Significantly, the radial velocity minimum is not coincident with
the velocity span minimum derived from their line bisector analysis.

The star $\eta$ Cep shows $>$99\% significant periodicities at 
$P=$164 days and $\sim$10 yrs. Walker et al. have speculated that
the 164 day periodicity was due to stellar rotation. No periodicities 
are detected in line asymmetry to $\sim$19 m/s by Gray (1994) with
measurements spanning four years. However we find the best fit radial
velocity amplitude at each of these periods is only 16 and 13 m/s 
respectively. If a direct correlation between a radial velocity 
measurement and a line bisector measurement exists, such signals would
be below his detection limit. By analogue with $\epsilon$ Eri, we
speculate that the 10 year periodicity in $\eta$ Cep might be 
linked to a magnetic cycle, but we cannot be certain of its origin.

Other marginal periodicities appear in the data for HR 8832 and 
$\theta$ UMa. Again by analogue with other stars, in this
case $\epsilon$ Eri and $\eta$ Cep, we might speculate that these
shorter periodicities are due to stellar rotation, however no 
certainty can be attached to their origin. 

\begin{figure}
\plotone{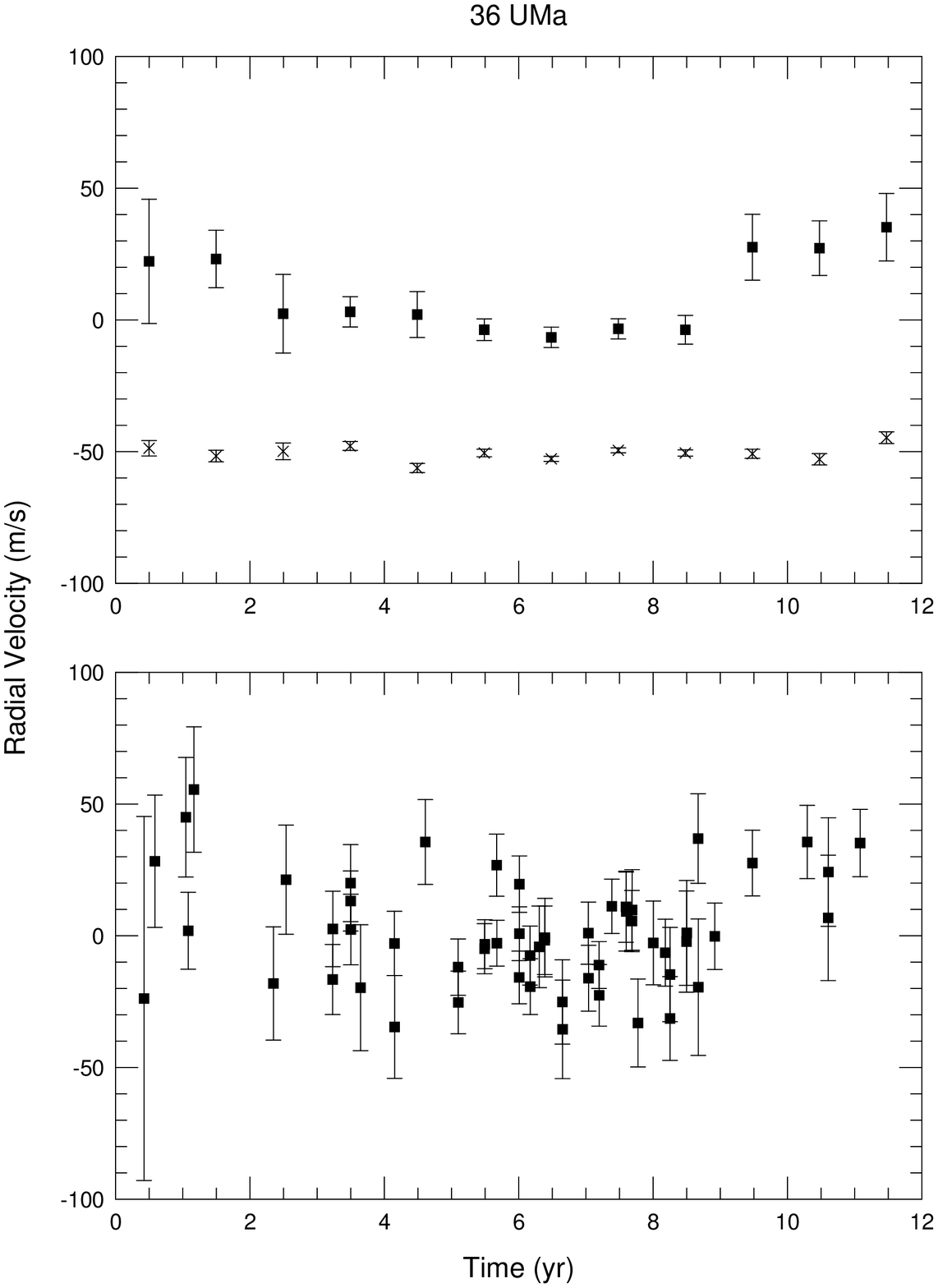}
\caption{\label{star-rv}
Radial velocity data for the stars (a) 36 UMa and (b) $\beta$
Vir, for which long term trends are observed using both our analysis
technique and the Walker et al. analysis. Both the published velocity 
data and the weighted mean of the velocities for each year are shown.
The binned data for the sample of 14 stars taken together is shown offset 
by $-50$ m/s. No trend similar to that found in either 36 UMa or $\beta$
Vir is present, indicating that a systematic error common to all stars
is unlikely.  Errors are taken from Walker et al. 1995, while the errors 
in the binned points are derived from the central limit theorem 
1/$\sqrt{n}$ improvement in the error of a multiply sampled mean.}
\end{figure}

We also find that two stars in the subset (36~UMa and $\beta$~Vir)
show best fit minimum masses which, for fitted periods longer than 
$\sim$12 years, rise above the curve for which the best fits are 
random with 99\% and 99.9\% probability. Walker et al. find similar 
trends in these stars but make no firm conclusions based upon their
analysis.  Are these signals indications of long period companions, 
or are they also due to stellar effects? We show the data for these 
stars in figure \ref{star-rv}, both raw and binned by year. While the 
raw data show no obvious signals, the binned data, particularly for 
36~UMa, show some indication of a partially complete sinusoid. We 
note that in both cases, the curvature in the velocity trends is of 
the same sign and the portion of the sinusoid is similar, which
suggests a long term calibration error.  However, the binned data for 
all stars taken together shows no such trend so a systematic 
explanation seems less likely.

\subsection{A Check by Monte Carlo Analysis}

Each of the stars in the Walker et al. sample average 3-5 measurements 
per year over the 12 year period and, according to the results of section 
\ref{sparse}, this number should be sufficient for application of 
our analytic apparatus. We note however, that implicit in our analytical
derivation of mass limits is the assumption that the measurements be at
least somewhat regularly distributed. In the case of the Walker et al.
data, this is not always the case. The data are irregular on both short
time scales (ie. 3 night runs consisting of 1--3 velocity measurements
per star per run) and longer time scales, for which more data may be
loosely clustered on several year time scales due to changes in 
observing procedures etc. Because of these irregular sampling patterns
fits may be less tightly constrained than a more evenly spaced data 
set, and the limits provided by our analytic apparatus may become 
misleading.

Since the Walker et al. data are rather irregular, we examine the effect
on our analysis technique by performing a Monte Carlo experiment and 
comparing the result to our analytic formalism. We create synthetic data
sets using a constant value of the error, $\sigma_p$, equal to the rms
scatter of the observed velocity measurements for each star. This value
is input into eqn. \ref{vrand-eq} to derive individual simulated velocity
measurements. We use the observation times given by the data itself. 
We fit the measurements and derive best fit velocity amplitudes (and
corresponding $M_c\sin(i)$ values) for periods between 3 days and 100 years
Each synthetic datum used in the fit is weighted with the internal error 
in that point (quoted by Walker et al.) as $1/\sigma_i(t_i)^2$.

The results of these experiments are shown as the solid histograms
in figure \ref{datamc-bf}. In general, the agreement between
the analytic limits and the Monte Carlo experiments is good. However
a small systematic trend towards larger limits for the Monte Carlo
experiments is found. Typically the difference is 10\% or less,
however, in the most extreme case ($\theta$~UMa), the limits produced 
are about 20-30\% higher than with the analytic method. This star has
the shortest time baseline of any in the sample as well as one of the
largest degrees of clumping of any star in the sample, as measured
by the ratio of the number of data to the number of runs. A test in 
which we replace in eqn. \ref{vprob-eq-long} the number of data, $n_0$ 
with the number of runs recovers the Monte Carlo results for this
star quite well. The star $\epsilon$~Eri, for which the data are 
the most highly clumped of any star in the sample, also produces 
analytic limits lower than reproduced with the Monte Carlo experiment. 
In this case, replacing $n_0$ with the number of runs produces limits
much larger than the Monte Carlo result, so we cannot recommend such
a procedure for general use.

In the case of one star (HR 8832), two measurements were made with
a 3-1/2 year separation from any other measurement for that star.
This case provides an interesting test in the limit of very 
irregularly spaced data.  We find that the limits derived from the
Monte Carlo experiment are quite similar to the analytic result
except in the range between about 8 and 12 years, where limits
some 20\% larger than those derived via eqn. \ref{m-eq} are found. 
The longer period fall-off characteristics are unaffected by the
irregularity.

\begin{figure}
\plotone{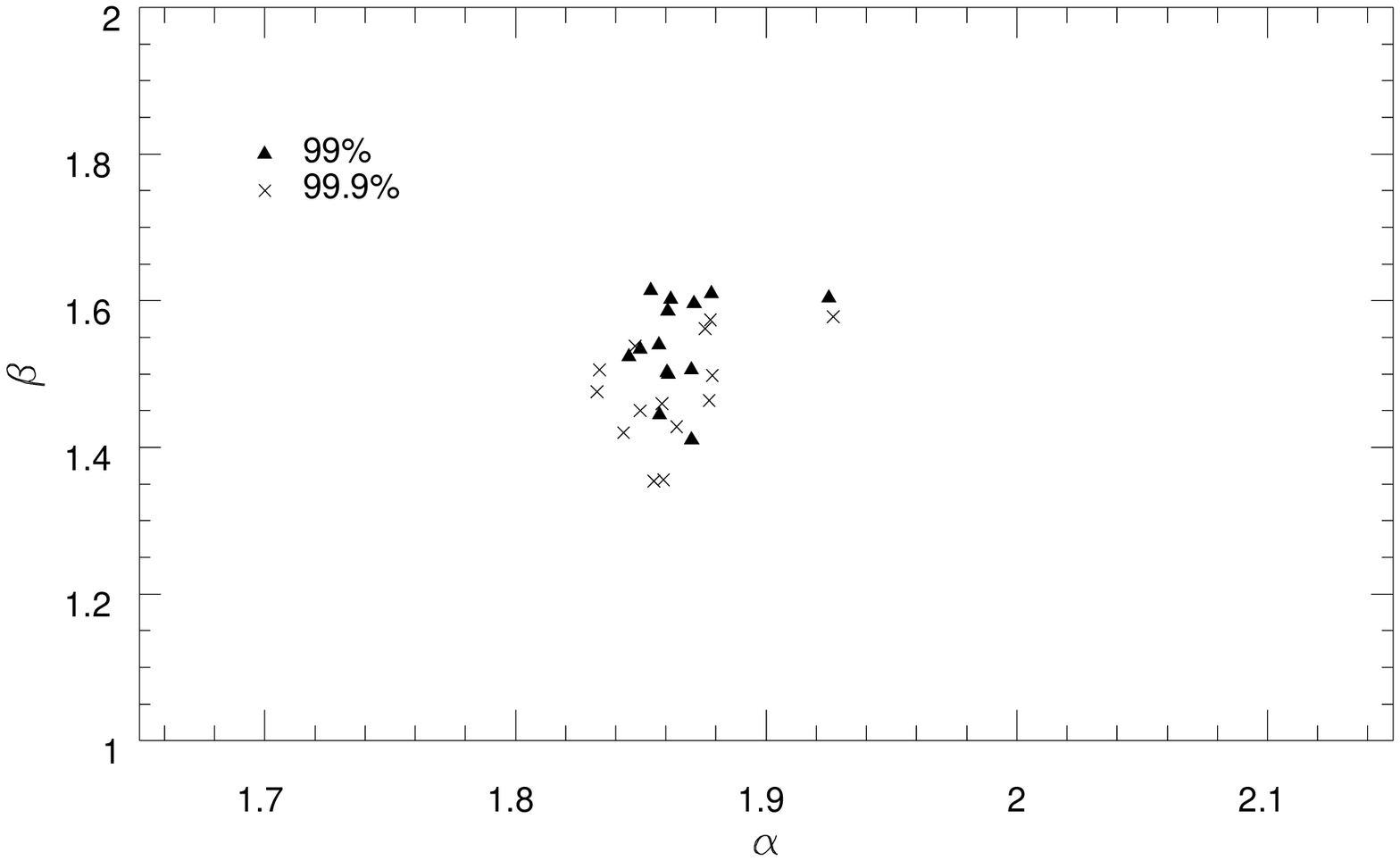}
\caption{\label{datal-bf}
Best fit values for the long period sensitivity fall off parameters
$\alpha$ and $\beta$ derived from the data. Values for both 99\% 
(triangles) and 99.9\% ($\times$'s) fall-off are shown. The two
points lying to the right of the main group originate from the star
$\theta$ UMa and are consistent with the monte carlo results for
a 6 year data span.}
\end{figure}

The sensitivity fall off at long periods for each star is similar to 
that produced in the synthetic data. We show the derived fit values for 
the power law exponent and the long period turn off in figure 
\ref{datal-bf} for the sample of 14 stars. The sharp upturn in limiting 
velocity at long periods produces a power law exponent which is 
best fit with values near $\alpha=1.86$, while the turn-off period, $\beta$,
is best fit with values near $\beta=1.45$, but with a larger scatter 
than is present in the synthetic data. Because the scatter is by its
nature rather unpredictable from star to star, we retain the low 
$\beta=1.3$ value found for synthetic data when determining limits
via eqn. \ref{vprob-eq-long} or \ref{m-eq}. 

The limits provided by our analytic expression produce upper bounds
which are ordinarily $\lesssim$10\% different than those produced via 
Monte Carlo experiments. In the most irregularly spaced data, 
a difference of up to 20-30\% can occasionally be produced. In several
cases, the difference results in possibly spurious `detections' of 
marginal signals by the analytic technique where the Monte Carlo limits
do not show that the periodicities are significant. In some of these
cases we are able to attribute the detections to physical processes
discussed in the literature. In no case do the analytic limits exceed 
the 99.9\% level of probability where the Monte Carlo result did not 
also show at least a 99\% probability. Despite this level of difference,
our conclusions about the significance or lack of it for any periods 
and companion masses for each star in the Walker et al. data remain 
unchanged. We are confident that this method can be relied upon to
obtain probabilities that a given set of data contains a periodic
signal.

\subsection{Sensitivity to Short Period Planets}\label{limits}

For the star with the lowest companion mass limits
($\epsilon$ Eri), we have also plotted in figure \ref{datamc-bf} 
several recent planet detections (Mayor and Queloz 1995, Marcy 
and Butler 1996, Butler and Marcy 1996, Latham et al. 1989, 
Gatewood 1996, Noyes et al. 1997) and Jupiter. Extra solar
planets with combinations of period and mass like those shown
would have been readily detected by Walker's radial velocity
measurements. These stars do not have such companions.  For most
of the stars in the sample, the data are too noisy to have 
reliably detected a radial velocity signature such as would
be predicted for the companion to Lalande 21185, announced by
Gatewood (1996) but which remains unconfirmed.  Planets such as
Jupiter, which would appear at $P=12$ years with a typical value
of $M_c\sin(i)$=0.64$M_J$ would not have been reliably detected.
The best fit values exceed this period/mass combination in 40\%
of the sample. 

The analysis of Walker et al. sets upper limits to the mass 
of companions in their sample ($\times \sin(i)$) of $\leq 1 M_J$ 
and $\leq 3 M_J$ in periods of less than 1 year and 15 years 
respectively. In general, our analysis provides limits which are 
somewhat lower than theirs in both long and short period orbits.
For one year periods, we can limit companion $M\sin(i)$ values
to $\leq 0.7 M_J$ for all but three stars in our subset and
$\leq 1.0 M_J$ for the rest. In 15 year orbits, our analysis
limits possible $M\sin(i)$ values to $\leq 1.5 M_J$ for every star
except $\theta$ UMa, for which only 6 years of data were gathered.
For this star, the limit is $\leq 4.0 M_J$.  We have also extended
range over which companion signatures are constrained to shorter
periods than were analyzed in Walker et al. The limits for these
extreme short period orbits ($P<40$ days) correspond to companion
masses ($\times \sin(i)$) below 0.4~$M_J\sin(i)$. 

Under either our own analysis or the original analysis of Walker 
et al., the companion mass limits derived from the data essentially
eliminate brown dwarfs and large Jovian planets with periods
$\lesssim 15$ years, barring very unfortunate inclinations. 
Given the detections of significant periodicities by either our  
analytic treatment or Monte Carlo experiments, we find more signals 
present than can be attributed to purely random data. In some cases,
such detections may be due to physical mechanisms other than a 
companion, and we have compared these to known periodicities due
to stellar rotation or magnetic cycles, where they have been 
identified in the literature. In no case do the limits eliminate 
the possibility of gas giants such as exist in our solar system
or low mass brown dwarfs, especially in the period/radius range
$\geq$12~yr/5 AU where theory predicts such companions.  

\section{Strategies for Large Radial Velocity Surveys}\label{strategy}

There are currently six active groups with programs for radial 
velocity searches at the $<$10-20 m/s level. Three groups began
searches at this precision in 1987-88 (Cochran \& Hatzes 1994
(Texas), McMillan et~al. 1994 (Arizona), Marcy \& Butler 1992
(Lick)), while one (Duquennoy \& Mayor 1991 (Geneva)) have used
lower precision measurements with the CORAVEL spectrometer 
($\sim300$ m/s) to investigate stellar binary companions and
have recently built a new spectrometer (ELODIE) to allow 
$\lesssim$15 m/s precision measurements to be made.  The latest 
high precision searches (K\"urstner et~al. 1994 (ESO), Brown 
et~al. 1994 (CfA)) began in 1992 and 1995 with quoted precision
of $\sim$4-7 m/s and $\sim$10 m/s respectively.  Another group
(Walker et~al. 1995 (UBC)) concluded a 12 year search in 1992.
Two others (Mazeh et al. 1996 (CfA), Murdoch et~al. 1993 
(Mt John NZ)) obtain precision of $\sim$500 m/s and $\sim$60 m/s 
respectively.  

The recent discoveries of sub-stellar mass companions around other
stars have stirred new interest in very large radial velocity 
surveys.  The Geneva group for example, intends to expand their
search to $\sim$500 stars in the northern hemisphere (ELODIE)
and another $\sim$800 in the southern hemisphere (CORALIE), and
other groups have similar expansions underway. In order to observe
as many stars as possible with a finite telescope allocation, 
such large surveys must necessarily aim toward the most efficient
use of the available observing time. The goal of such large surveys
might be properly stated as ``What fraction of stars have a companion
(or a system of companions) and what is the distribution of the 
masses, periods and eccentricities of those companions?''. 

In order to answer this question three criteria must be met. First,
an observer must first detect a variation in the radial velocities 
measured for a star about which a prospective companion orbits. 
Second, the observer must determine the origin of such variations
by fitting a Keplerian orbit and by making additional photometric
or spectroscopic observations to constrain effects due to the 
stellar photosphere. Finally the observer must determine the extent
to which the survey is complete: what fraction of stars which were
observed may have companion signatures which went undetected over
the course of the survey? Based on the analysis in this paper, we
can suggest strategies for the most efficient methods of detecting
radial velocity signatures and which also provide meaningful upper
limits on the amplitudes of undetected signatures.

Let us suppose that the random error for each measurement is 
dominated by photon noise, ie. that $\sigma_p\propto 1/\sqrt{t}$,
where $t$ is the length of a single observation. This should be
the case provided detector read noise is not significant. It 
follows from eqn. \ref{vprob-eq} that the limiting amplitude,
$K$, is proportional to $1/\sqrt{t_n}$, ie. $K$ depends on the
total integration time devoted to a star, $t_n$, independent
of number of observations making up that time. In other words,
as long as eqn. \ref{vprob-eq} holds and the total integration
time is the same, making many lower precision measurements is
equivalent to making fewer high precision measurements. Since
constraining additional orbit parameters such as eccentricity
is at its most simplistic level an exercise in detecting higher
order Fourier components of the signal, this equivalence holds
for orbit determinations as well as for detection of a periodic
signal. We caution, however that with lower precision data,
larger amplitude systematic errors may go undetected. With 15 m/s
precision for example, the signal of Jupiter could be completely
obscured by a hidden systematic error of amplitude $\sim$10~m/s. 

\begin{figure}
\epsscale{.9}
\plotfiddle{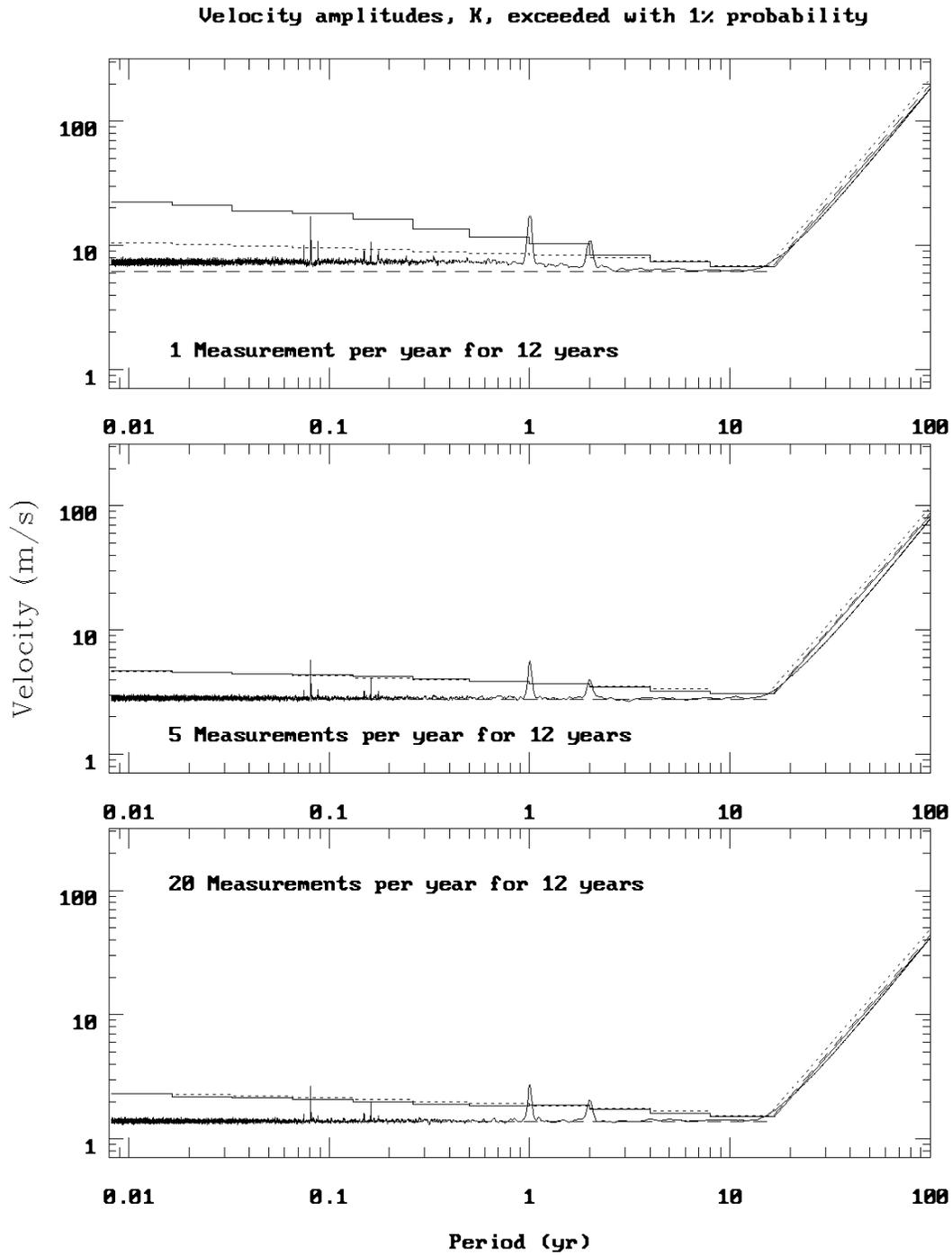}{6.79in}{0}{72}{72}{-230}{-20}
\caption{\label{mc-anal-cmp}
Monte Carlo (solid line) and analytic results (dashed line) for 
the velocity amplitude which, for a given fitted period, is 
exceeded by chance in 1\% of analyses of simulated random data. A
second solid line (histogram) shows the monte carlo results for
the amplitude which is exceeded at any period within a range of
approximately one octave, while the dotted histogram line shows
the result from our analytic expression, eqn \ref{vprob-eq}.
For each of the experiments in the bottom two frames (5 and 20 
obs./yr), the analytic and monte carlo results are indistinguishable.
Assumed windowing at the lunar and annual periods as well as their 
double period counterparts are excluded from the monte carlo limits 
in their respective ranges.}
\end{figure}

\begin{figure}
\plotfiddle{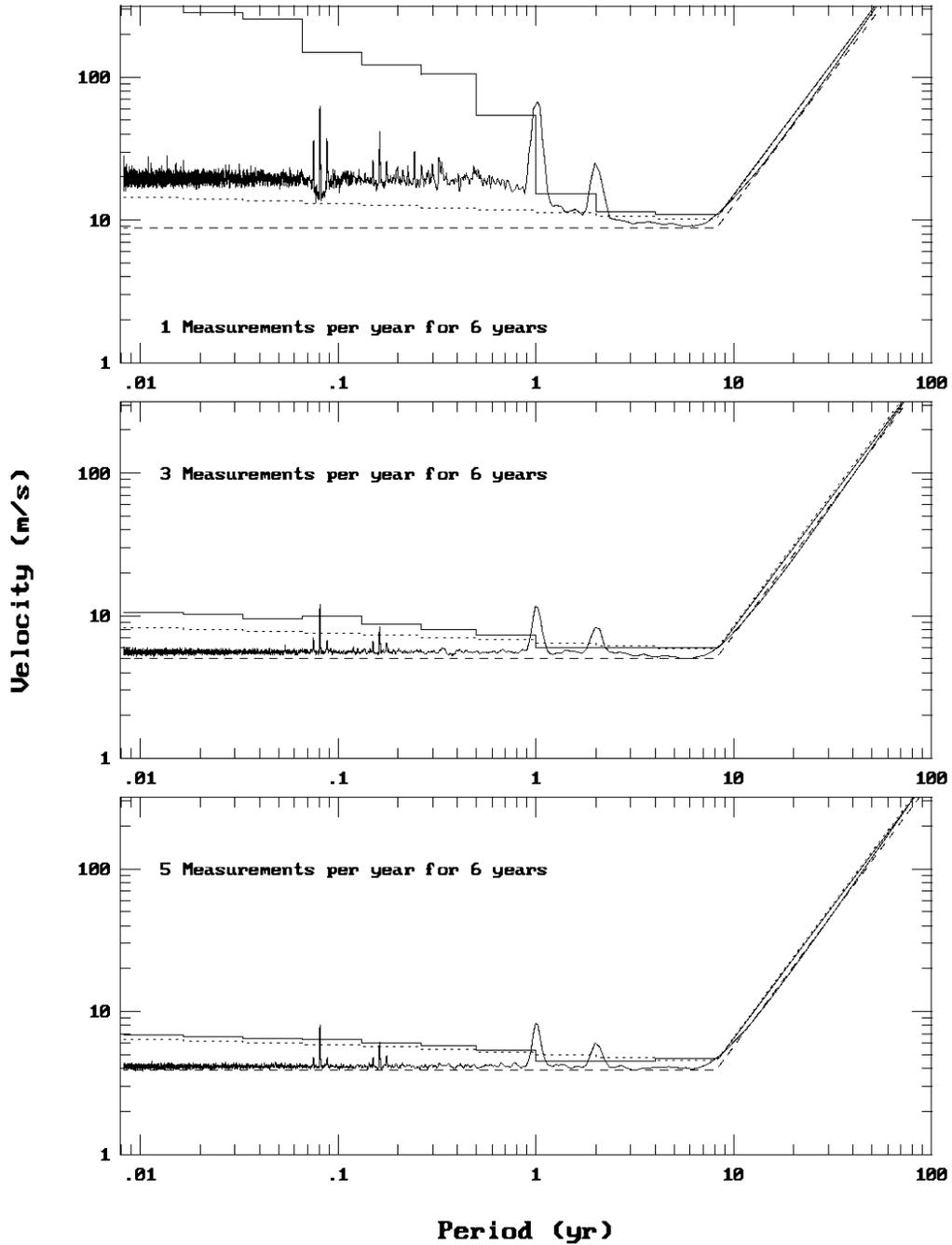}{6.79in}{0}{72}{72}{-230}{-20}
\caption{\label{vlim-sparse6}
Velocity limits given by monte carlo simulations (solid), 
single period ($N$=1) analytic limits (dashed). The histograms
represent the monte carlo (solid) and analytic (dotted) limits
for octave sized ranges. Each of the histogram limits ignore the
periods affected by the assumed lunar and annual windowing of 
the data. A 6 year window is assumed with 1 (top), 3 (middle) and
5 (lower) observations per year.}
\end{figure}

From section \ref{monte-carlo}, to insure than eqn. \ref{vprob-eq}
holds, the number of observations, $n_0$, must be at least $\sim$12 
and preferably as high as 20 in order to constrain octave period 
ranges.  The limits are degraded most severely for periods less 
than 1--2 years. Longer periods limits are nearly identical to the
analytic prediction even for very sparse data (see figures
\ref{mc-anal-cmp} and \ref{vlim-sparse6}). This is because 
with only a few observations of each star, a companion signature 
could still slip through undetected if by some unfortunate 
coincidence its radial velocity ``zero crossings'' corresponded 
to the times at which the star was observed. For $P\lesssim 1$ year,
there are very many independent periods, so that the possibility 
of any one of them coincidentally undergoing such a zero crossing
event is very high. For $P\gtrsim 1$ year, where there are relatively
few independent periods, such a condition becomes much more unlikely.

The cost in observing time to obtain useful limits if there are 
few observations is great. When the data are sparse and eqn.
\ref{vprob-eq} breaks down, for example with a total of either 
$n_0=6$ or $n_0=12$ observations and the same total integration
time, our Monte Carlo simulations show that the limiting amplitude
is in fact twice as big for any single period, and ten times as 
big for octave period ranges. Because 12 much lower precision
measurements would identically constrain short period signals as 6
high precision measurements, the increase in sensitivity translates
to a reduction factor of 4 or 100 in the amount of observing time
required to identically constrain the existence of companion 
for any single period or over octave ranges in orbits of
$\lesssim 2$ years. 

As an example of a strategy which addresses this concern, suppose 
a survey is to observe 500 stars and is to last at least 12 years.
Let us also suppose that 1/4 of the use of a telescope is dedicated 
to the radial velocity measurements, yielding about 400 hours of
integration per year to be divided among the stars in the survey. 
The total number of observations to obtain 12 for each star is 6000.
A good ``quick look'' could be obtained after the first two years if
each observations takes 2$\times$400/6000 hours = 8 minutes.

Butler et~al. (1996) report that precision of $\sim$3~m/s can be 
obtained in a 10 minute exposure of a magnitude $V=5$ star on a 
3 meter class telescope. However, in a large survey most stars will 
be dimmer than $V=5$, with a practical limiting magnitude between
$V=7$ and $V=8$, depending on the size of the survey. If the average
star is of magnitude $V=7$, a measurement with 3 m/s precision
would nominally require about 60 minutes.  With such long duration
measurements each star in the program would average less than one 
observation per year and would make a large, high precision survey 
unfeasible. In order to complete a large survey at $\leq$5 m/s
precision a large allocation of time on an 8--10 m class telescope
would be required. If instead we allow reduced precision measurements
of $\pm$10~m/s, using our assumption that the precision is
proportional to $1/\sqrt{t}$, a single measurement would require only 
about 6 minutes on a 3~meter telescope, which would be feasible for 
a large survey.

For this to be practical without poor observing efficiency, the 
time from the end of an observation to the beginning of the next
on a new star must be short, ideally a minute or less.  During this
time, the telescope must be slewed to the new star, while the 
CCD with the exposed spectrum is read out.  Automatic slewing 
and acquisition would make this quite practical.  Also, for a
typical spectroscopic CCD with around 3 million pixels, the 
required read rate of 100 kpixel/sec should be readily achievable
at negligible read noise with current devices. A 8 minute cycle
time with 6 minutes data acquisition would thus be a reasonable
target, and yield 72 minutes of integration for each star, spread
through the first two observing seasons.

With this strategy, an observing program should be able to sustain
6 measurements of every star every year that the
program is continued. If after two years, variations are detected
in some stars, additional measurements of those stars would
be possible if constant velocity stars were observed only 1--3 
times per year. This compromise has the advantage that limits on
companion masses are well constrained by such density of points
and orbital solutions, should a star's velocity later be observed
to vary, would also be well constrained. A second advantage is that
after the first two years, strong limits on the existence of a 
companion signal are available for short periods and these limits
extend to longer periods incrementally as long as the program is
maintained.  In contrast, a high precision/sparse observation 
strategy with say 1-2 measurements per year, will strongly limit
short period signatures only after 6 or more years of the program
has passed.

\begin{figure}
\plotone{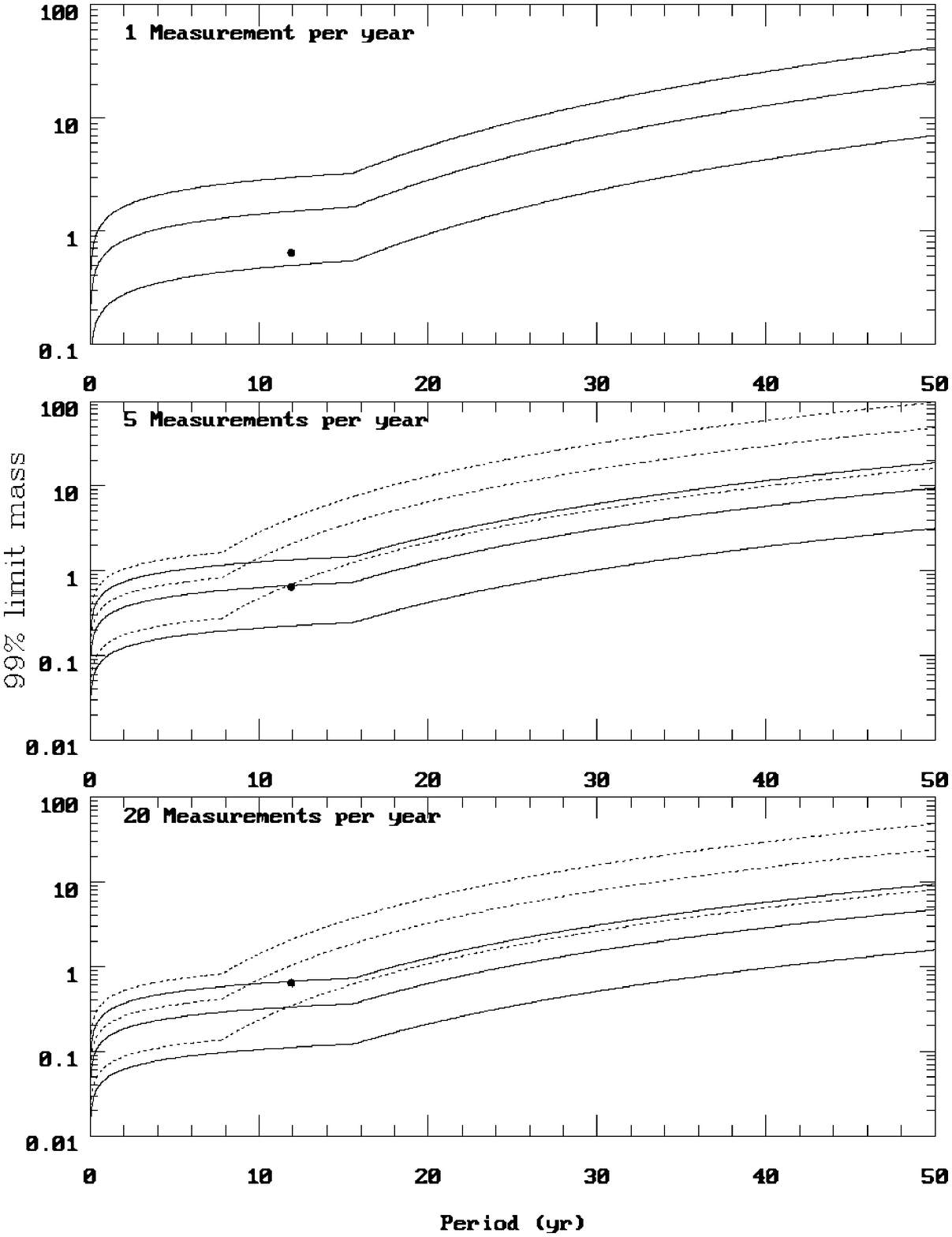}
\caption{\label{mlim-anl}
Limits above which data of a given quality and duration 
constrain the mass of companions at the 99\% level for any single 
period. Solid lines represent a 12 year span of data taken with 5, 15 
and 30 m/s precision while dotted lines represent a comparable 6 year
span.  Limits for the 6 year baseline with one only measurement per 
year are omitted here because they do not correspond to results from 
monte carlo experiments (see section \ref{sparse}).}
\end{figure}

As a second example of observing strategy, we consider a search 
for a Jupiter mass planet with the same 12 year period as
Jupiter, around a star with the same mass as the sun, what 
accuracy measurements are needed over what period to ensure only
1\% probability of a false detection for any given star? For sets
of observations spanning 6 and 12 years figure \ref{mlim-anl} 
shows the limiting mass above which a companion would be detected
with 99\% probability in a given set of data at any single period.
A 1~$M_J$ companion in a 12 year orbit around a solar twin will 
have a best fit mass of $\sim 0.64 M_J$ assuming a random 
inclination. We require via eqn. \ref{m-eq}, that data be taken
with precision $\pm$5 m/s for 12 years with a single observation
per year in order to detect such a companion. Increasing to 5 or
20 observations per year, only 15 or 30 m/s are required, 
respectively, with the requirement that no hidden systematic errors
are also present in the data. For an identical number of 
measurements per year, a 6 year baseline requires more than 6 
times the precision in each measurement to similarly constrain 
a long period companion, or 36 times the observing time each 
year. Clearly the cost of impatience is very high.

\acknowledgments
We would like to thank the referee, Paul Butler, for helpful comments
and criticisms in his referee's report. Alan Irwin also provided 
helpful criticism of this manuscript. We are grateful to several
people for providing the radial velocity data used in this work.
Among them are Gordon Walker and Alan Irwin for providing a portion of
their radial velocity data prior to publication, Michel Mayor for
sharing the data for 51~Peg and HD~110833 and Geoff Marcy for sharing
an electronic copy for a second, independent data set for 51 Peg.  
M. Mayor provided commentary on the results of our analysis and 
a discussion with Didier Queloz outlined reasons for which the data
for HD~110833 may have contained systematic errors in several 
measurements. W. Benz provided impetus for the section on sensitivity
to large signals and Bob McMillan, Adam Burrows, Paul Harding and Heather
Morrison provided other helpful discussion. This work was partially
supported under NASA Grants NAGW-3406 and NAS7-1260.

\newpage

\begin{deluxetable}{rrrcccccc}
\tablecolumns{8}
\tablecaption{\label{star-tab} The Subset of 14 Stars Included in our Analysis}
\tablehead{
\colhead{HR}  & \colhead{HD} & \colhead{Name} & \colhead{M/M$_\odot$}
& \colhead{$\sigma_i$ } & \colhead{ $\sigma_p$ } 
& \colhead{Number }     & \colhead{Number }  & \colhead{Duration } \\
\colhead{}    & \colhead{}   & \colhead{}     & \colhead{}
& \colhead{(m/s)}       & \colhead{(m/s)}
& \colhead{Obs.} & \colhead{Runs} & \colhead{ (yr)}
}

\startdata
 509 &  10700 & $\tau$ Cet       & 0.87 & 13 & 17 & 68 & 39 & 11.7 \nl
 937 &  19373 & $\iota$ Per      & 1.15 & 15 & 18 & 46 & 29 & 10.8 \nl
 996 &  20630 & $\kappa^1$ Cet   & 0.98 & 13 & 20 & 34 & 22 & 10.0 \nl
1084 &  22049 & $\epsilon$ Eri   & 0.82 & 14 & 16 & 65 & 34 & 11.1 \nl
1325 &  26965 & o$^2$ Eri        & 0.84 & 14 & 19 & 42 & 28 & 11.0 \nl
3775 &  82328 & $\theta$ UMa     & 1.45 & 24 & 21 & 43 & 23 & \phn6.0 \nl
4112 &  90839 & 36 UMa           & 1.08 & 16 & 21 & 56 & 36 & 10.7 \nl
4540 & 102870 & $\beta$ Vir      & 1.22 & 14 & 26 & 74 & 48 & 11.7 \nl
4983 & 114710 & $\beta$ Com      & 1.09 & 16 & 18 & 57 & 40 & 11.4 \nl
5019 & 115617 & 61 Vir           & 0.98 & 13 & 18 & 53 & 35 & 11.4 \nl
7462 & 185144 & $\sigma$ Dra     & 0.85 & 13 & 19 & 56 & 37 & 11.5 \nl
7602 & 188512 & $\beta$ Aql      & 1.30 & 12 & 14 & 59 & 39 & 11.4 \nl
7957 & 198149 & $\eta$ Cep       & 1.36 & 12 & 19 & 58 & 39 & 11.2 \nl
8832 & 219134 & \nodata          & 0.79 & 11 & 15 & 32 & 23 & 10.6 \nl
\enddata
\end{deluxetable}

\newpage

\begin{deluxetable}{c}
\tablecolumns{1}
\tablewidth{1in}
\tablecaption{\label{range-tab} Period Ranges}
\tablehead{
}
\startdata
3-6d  \nl
6-12d  \nl
12-24d  \nl
24-48d  \nl
48-96d  \nl
96d-0.5yr  \nl
0.5-1yr  \nl
1-2 yr  \nl
2-4 yr  \nl
4-8 yr  \nl
8-12 yr  \nl
$>$12 yr  \nl
\enddata
\end{deluxetable}

\begin{deluxetable}{rrcccc}
\tablecolumns{6}
\tablecaption{\label{signif-tab} Detection of Significant 
Periodicities\tablenotemark{a}}
\tablehead{
\colhead{Period range}  & \colhead{Star} & \colhead{Period} &
\colhead{Companion\tablenotemark{b}}  &
               \multicolumn{2}{c}{Probability of Chance} \\
\colhead{}  & \colhead{} & \colhead{} & 
\colhead{Mass (M$_{\rm J}$)}  &
	       \multicolumn{2}{c}{Detection in Period Range} \\
\colhead{}   &  \colhead{}   &  \colhead{}   & 
\colhead{}   & \colhead{Analytic}  & \colhead{Monte Carlo} 
}
\startdata
3-6d      &  \nodata        &          &                 &          &          \nl
6-12d     &  $\epsilon$ Eri &    11.9d &   0.14          & $<$1\%   & \nodata  \nl
12-24d    &  \nodata        &          &                 &          &          \nl
24-48d    &  \nodata        &          &                 &          &          \nl
48-96d    &  $\epsilon$ Eri &    52.5d &   0.24          & $<$1\%   & \nodata  \nl
96d-0.5yr &  $\eta$ Cep     &     164d &   0.54          & $<$0.1\% & $<$1\%   \nl
          &  HR 8832        &     165d &   0.35          & $<$1\%   & \nodata  \nl
          &  $\theta$ UMa   &     179d &   0.63          & $<$1\%   & \nodata  \nl
0.5-1yr   &  \nodata        &          &                 &          &          \nl
1-2 yr    &  \nodata        &          &                 &          &          \nl
2-4 yr    &  \nodata        &          &                 &          &          \nl
4-8 yr    &  $\epsilon$ Eri &  $>$7 yr &   0.7\phn       & $<$1\%   & $<$1\%   \nl
8-12 yr   &  $\epsilon$ Eri &    10 yr &   0.95          & $<$0.1\% & $<$1\%   \nl
          &  $\eta$ Cep     &    10 yr &   1.2\phn       & $<$1\%   & $<$1\%   \nl
          &  $\beta$ Com    &    10 yr &   1.05          & $<$1\%   & \nodata  \nl
          &  36 UMa         &    10 yr &   1.1\phn       & $<$0.1\% & $<$0.1\% \nl
$>$12 yr  &  36 UMa         &    15 yr &   2.0\phn       & $<$0.1\% & $<$0.1\% \nl
          &                 &    25 yr &   5.3\phn       & $<$0.1\% & $<$0.1\% \nl
          &                 &    50 yr &  24\phd\phn\phn & $<$0.1\% & $<$0.1\% \nl
          &  $\beta$ Vir    &    15 yr &   1.9\phn       & $<$0.1\% & $<$0.1\% \nl
          &                 &    25 yr &   5.0\phn       & $<$0.1\% & $<$1\%   \nl
          &                 &    50 yr &  23\phd\phn\phn & $<$1\%   & $<$1\%   \nl
\enddata
\tablenotetext{a}{We exclude `significant' periods which coincide with
known annual or lunar windowing periods}
\tablenotetext{b}{Best fit mass assuming that the periodicity is
actually due to a companion.}
\end{deluxetable}

\newpage

\clearpage

\clearpage

\clearpage

\clearpage

\clearpage

\clearpage

\clearpage

\clearpage

\begin{figure}
\figurenum{\ref{datamc-bf}--cont}
\plotfiddle{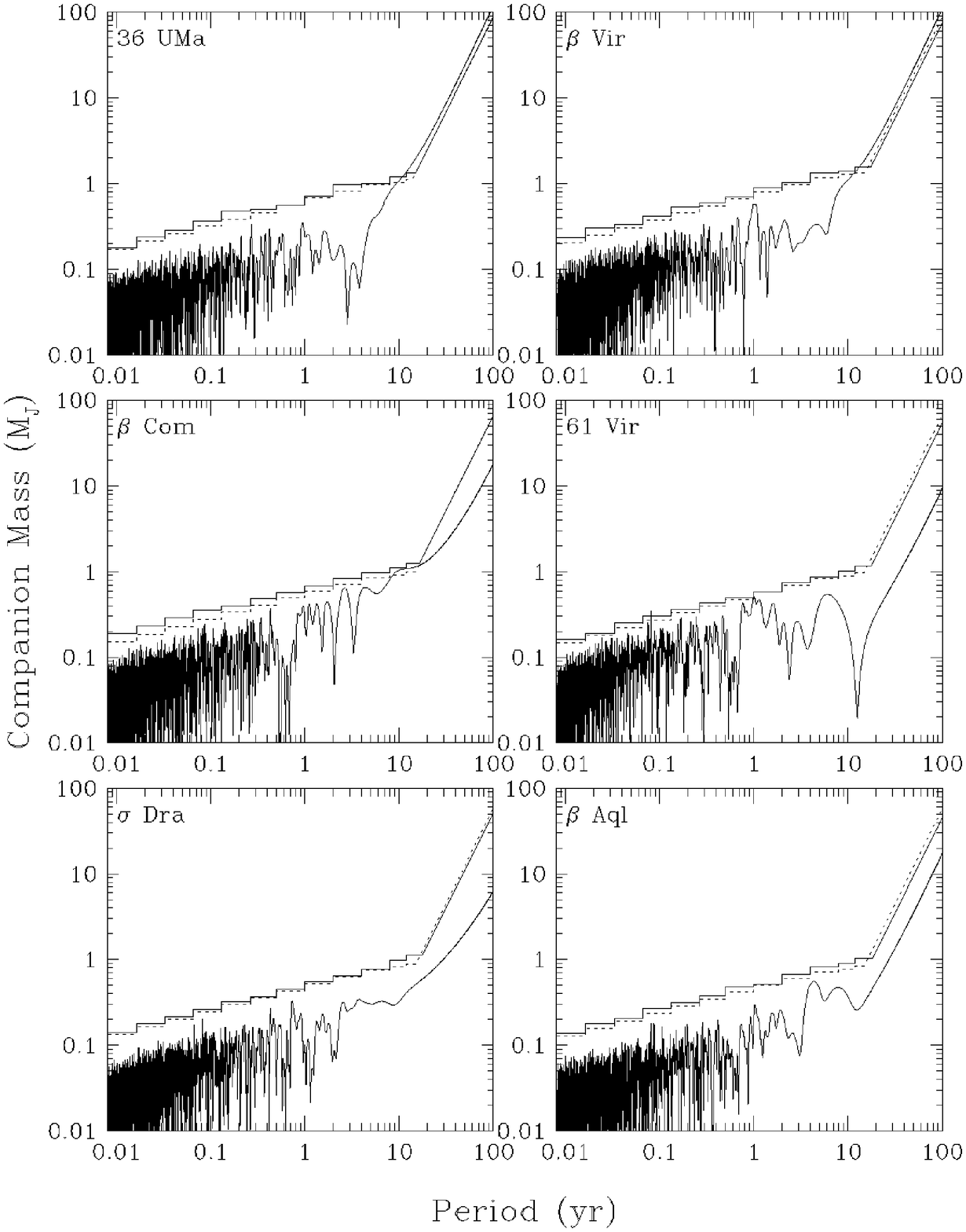}{7.2in}{0}{83}{83}{-260}{-40}
\caption{}
\end{figure}

\clearpage

\begin{figure}
\figurenum{\ref{datamc-bf}--cont}
\plotfiddle{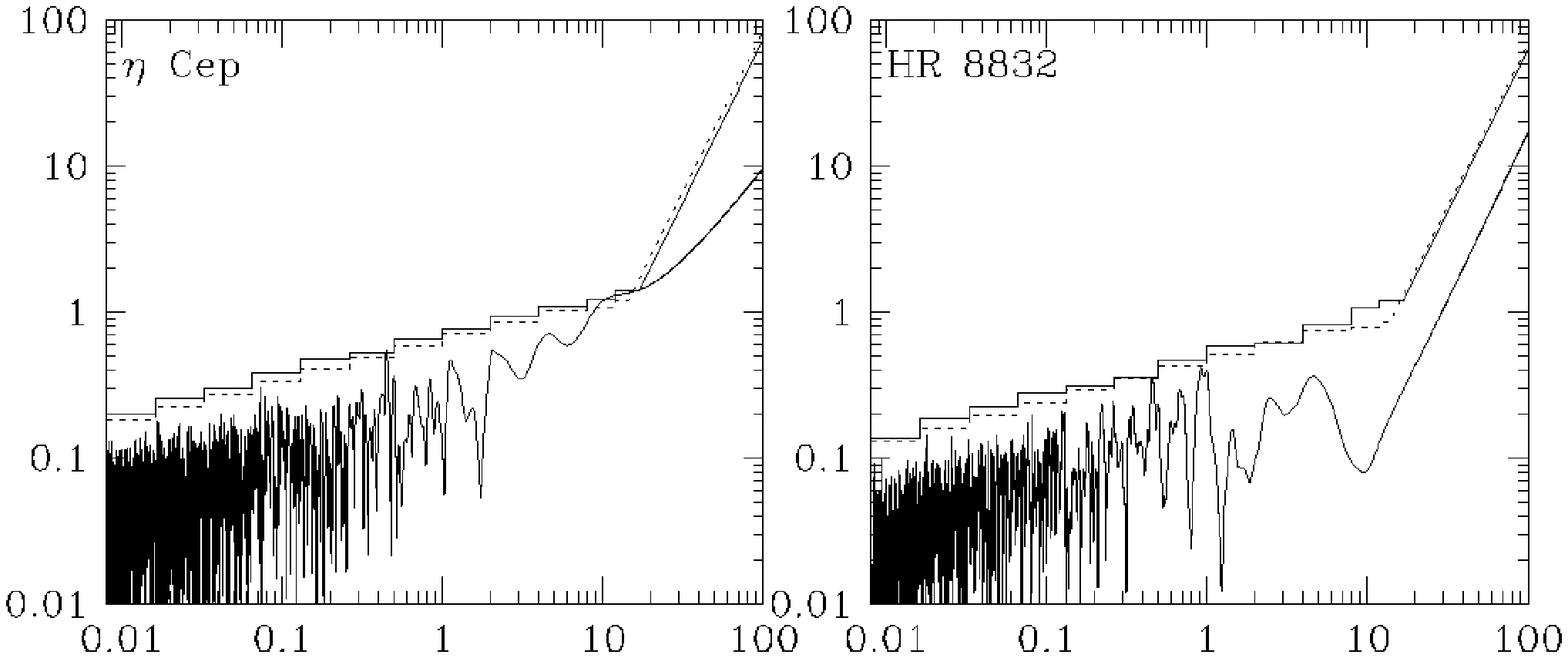}{7.2in}{0}{83}{83}{-260}{-190}
\caption{}
\end{figure}

\clearpage

\clearpage

\clearpage

\begin{figure}
\plotone{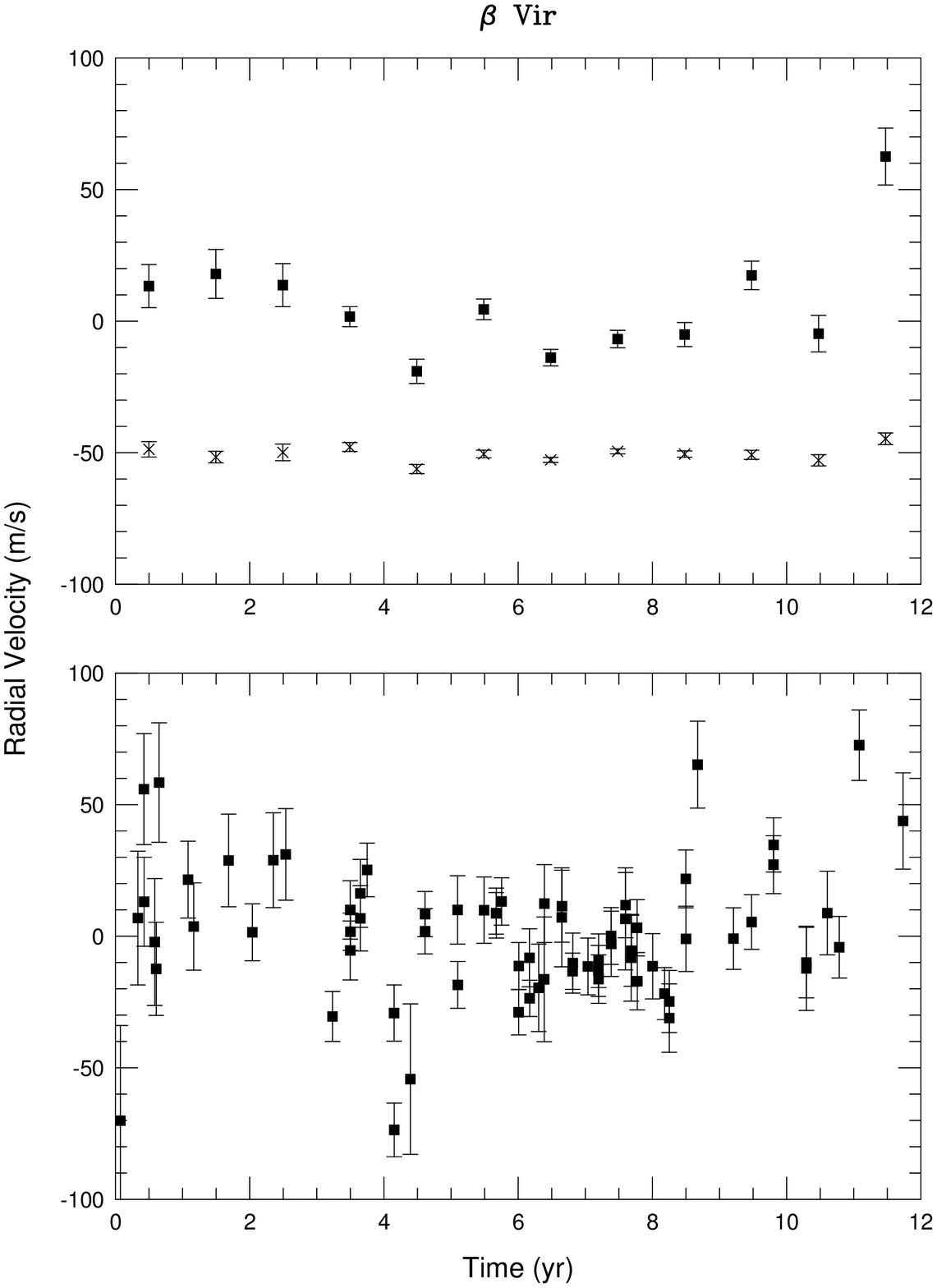}
\figurenum{\ref{star-rv}b}
\caption{}
\end{figure}

\clearpage

\end{document}